\documentclass[aps,10pt,a4paper,twocolumn,footinbib, superscriptaddress, floatfix]{revtex4-2}
\usepackage{graphicx}
\usepackage{amsmath}
\usepackage{amsfonts}
\usepackage{amssymb}
\usepackage{dsfont}
\usepackage{url}
\usepackage{hyperref}
\hypersetup{colorlinks=true, citecolor=blue, urlcolor=blue, linkcolor=blue}
\usepackage{amstext}
\usepackage[caption=false]{subfig}
\usepackage{epstopdf} 
\usepackage{mathtools} 
\usepackage{balance}
\usepackage{color,soul}
\usepackage{floatrow}
\usepackage{xr}
\usepackage[separate-uncertainty=true]{siunitx}
\DeclareSIUnit\sq{\ensuremath\Box}

\date{August 8, 2023}

\begin{document}

\title{Kinetic Inductive Electromechanical Transduction for Nanoscale Force Sensing}

\author{August K. Roos}
\thanks{These authors contributed equally.}
\author{Ermes Scarano}
\thanks{These authors contributed equally.}
\author{Elisabet K. Arvidsson}
\thanks{These authors contributed equally.}
\author{Erik Holmgren}
\author{David B. Haviland}
\email{haviland@kth.se}
\affiliation{Department of Applied Physics, KTH Royal Institute of Technology, Hannes Alfvéns väg 12, SE-114 21 Stockholm, Sweden}

\begin{abstract}
We use the principles of cavity optomechanics to design a resonant mechanical force sensor for atomic force microscopy. The sensor is based on a type of electromechanical coupling, dual to traditional capacitive coupling, whereby the motion of a cantilever induces surface strain that causes a change in the kinetic inductance of a superconducting nanowire. The cavity is realized by a compact microwave-plasma mode with an equivalent $LC$ circuit involving the kinetic inductance of the nanowire. The device is fully coplanar and we show how to transform the cavity impedance for optimal coupling to the transmission line and the following ampliﬁer. For the device presented here, we estimate the bare kinetic inductive mechano-electric coupling (KIMEC) rate $g_0 / 2\pi$ in the range \SIrange{3}{10}{\hertz}. We demonstrate phase-sensitive detection of cantilever motion using a multifrequency pumping and measurement scheme.
\end{abstract}

\maketitle

\section{Introduction} 

The atomic force microscope (AFM) is the most versatile and widely used form of scanning probe microscopy,imaging a wide variety of materials and basically any physical property that gives rise to a force between a sample and a probe \cite{Binning1986, Giessibl2003}. The key component of the AFM is the force sensor that produces the signal recorded while scanning the tip over a surface. Force transduction occurs in two steps: force on the tip is converted to motion of a micrometer-scale cantilever and cantilever motion is detected, typically by optoelectronic means. The response of the cantilever to force is greatly enhanced when operated in dynamic mode, oscillating at the frequency $\omega_m$ of a mechanical resonance with high quality factor $Q_m$.

Similarly, the detection of motion is also enhanced when the cantilever acts as a moving mirror in a high-Q resonant optical cavity. The fundamental limitations of the cavity optomechanical approach to motion sensing have been well studied, originally in the context of gravity-wave detection \cite{Braginsky1975} and later in a wide variety of micromechanical systems \cite{Aspelmeyer2014}, including AFM \cite{Rugar1991,Smith1995,Srinivasan2011, Halg2021}. The detected signal-to-noise ratio (SNR) increases with the circulating power in the cavity, until radiation pressure induces a backaction force that increases the motion noise of the cantilever. At the optimal power, cavity optomechanics can reach the standard quantum limit, where the backaction noise equals the measurement noise and the total uncertainty in the cantilever position is given by twice the quantum zero-point ﬂuctuations.

Cavity optomechanics is also realized with superconducting lumped elements where the ``cavity'' is an $LC$ circuit with a high-quality-factor resonance at $\omega_0=1/\sqrt{LC}$ in the microwave region. Electromechanical coupling has been explored in devices where motion of a wire modulates magnetic flux in a superconducting quantum interference device (SQUID), changing the Josephson inductance of a resonant circuit \cite{Nation2016, Rodrigues2019}. A more common type of coupling uses the electrostatic force on a flexible electrode of a capacitor \cite{Regal2008, Teufel2009}. SQUIDs have achieved rather large electromechanical coupling \cite{Zoepfl2020, Schmidt2020}, and the approach with capacitive coupling resulted in a milestone achievement: a resolved-sideband device that cooled a micromechanical mode with resonance frequency $\simeq$ \SI{10}{\mega\hertz} to its quantum ground state \cite{Teufel2011}. Numerous measurement schemes of fundamental interest and proof-of-concept experiments have been studied with circuit electro-mechanics \cite{Massel2011, Suh2013, Yuan2015, Pirkkalainen2015, Kalaee2019, Peterson2019, Bothner2020, Arnold2020} but the devices are generally ``self-sensing'' and not practical for scanning probe microscopy.

We report here on a cantilever device with an integrated motion detector based on electromechanical coupling. Cantilever motion induces surface strain that changes the kinetic inductance $L_k$ of a superconducting nanowire, detected as a shift in the resonance frequency $\omega_0 = 1/\sqrt{L_k C}$ of a microwave circuit. This coupling is particularly appropriate for AFM force transduction as it exploits the mechanical advantage of the cantilever (a mechanical lever), converting a weak force on the tip at the free end to concentrated strain at the ﬁxed end. To emphasize its dual nature and to stress that the supercurrent carries both mass and charge, we introduce the term kinetic inductive mechanoelectric coupling (KIMEC). KIMEC allows for a very compact microwave design with simple planar nanofabrication and easy integration of a sharp tip, making it highly suitable for AFM. While our device is reminiscent of the common resistive-strain sensor, KIMEC gives a reactive electromechanical transduction, which, in principle, is without dissipation and therefore intrinsically noise free.

\begin{figure*}[ht]
\includegraphics[width=17.2cm]{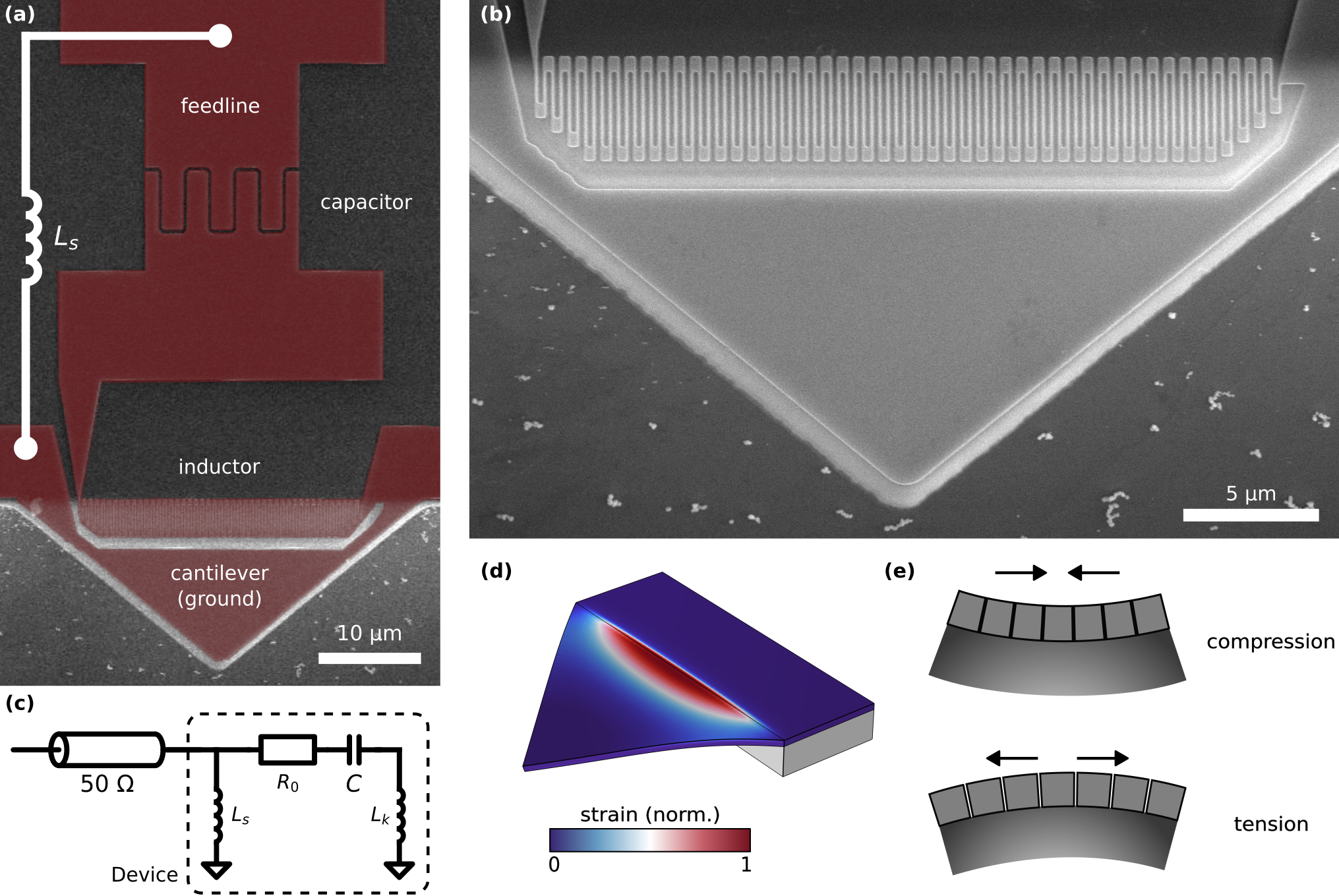}%
\caption{Electromechanical transduction via strain. (a) A false-color scanning-electron-microscope (SEM) image of the device showing the superconducting thin film of Nb-Ti-N (red) on a Si-N layer and Si substrate (gray background). The Si support is etched away under the triangular cantilever. The on-chip shunt inductance $L_{s}$ (illustrated) is either a wire bond or a short nanowire kinetic inductor. (b) A SEM image detailing the kinetic inductor, a superconducting nanowire meandering along the clamping line of the cantilever. (c) The equivalent circuit with the microwave mode modeled as a series $RLC$ circuit connected to a 50-$\Omega$ transmission line in parallel with a shunt inductance $L_{s}$. (d) A finite-element-method (FEM) simulation, showing the normalized volumetric strain of the first flexural mode, where red indicates tensile strain. The mechanical displacement is exaggerated for clarity. The meandering nanowire (not shown) is positioned in the region of maximum strain. (e) An illustration of a possible working principle of KIMEC, where strain compresses (stretches) grain boundaries, influencing Josephson tunneling.}%
\label{fig:device_design}
\end{figure*}

\section{Kinetic inductance and strain} 
Superconductivity, as explained by Bardeen-Cooper-Schrieffer (BCS) theory, results from electromechanical coupling. The electron-phonon interaction is responsible for the formation of Cooper pairs or $2e$-charged Bose particles that reside in a condensed ground state at the gap energy $\Delta_0$ below the Fermi energy \cite{Tinkham2004}. A wire with cross-section area $A$ carrying a supercurrent $I_s = 2e n_s v A$, where $n_s$ is the Cooper-pair density moving at velocity $v$, also carries mechanical energy due to the Cooper-pair mass $2m_e$. The kinetic energy in a length of wire $\ell$ can be expressed in terms of an effective inductance $E_k = \frac{1}{2} (n_s A \ell) (2m_e) v^2 =\frac{1}{2} L_k I_s^2 $, or kinetic inductance per unit length:
\begin{equation}
   \frac{L_k}{\ell} = \frac{m_e}{2 e^ 2n_s A}.
\end{equation}
In nanowires with small $A$ made from ``dirty'' superconductors with small $n_s$, the kinetic inductance can far exceed the electromagnetic inductance per unit length $L/\ell \simeq \mu_0$. We realize a compact lumped-element superinductor at microwave frequencies by winding such a nanowire into a meandering structure [see Fig.~\ref{fig:device_design}(b)].

Within BCS theory, the pair density is given by \cite{Tinkham2004},
\begin{equation}
  n_s = \frac{N(0) \Delta_0}{2} \simeq N(0) \hbar \omega_D \exp \left( -\frac{1}{N(0)V} \right),
\end{equation}
where $V$ is the electron-phonon coupling energy, $N(0)$ is the density of states at the Fermi energy ($N(0)V \ll 1$) and $\omega_D = \pi a/ v_s$ is the Debye frequency, defined by the lattice constant $a$ and the speed of sound $v_s$. Tensile or compressive strain in the superconducting material affects these parameters, leading to a change of $L_k$. The ability to modulate the pair density with mechanical strain provides a mechanism for force transduction.

\begin{figure*}[ht]
\includegraphics[width=17.2cm]{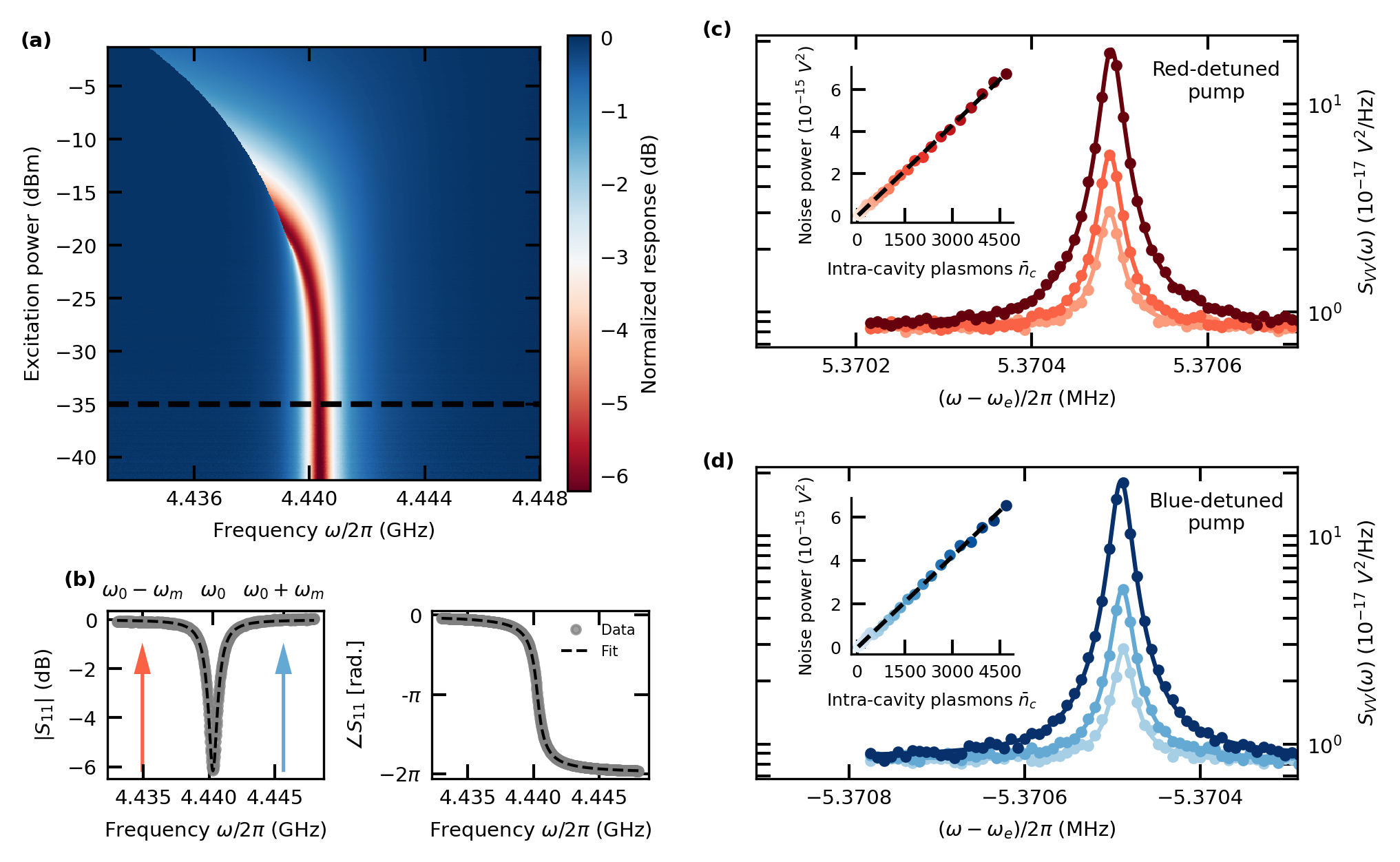}%
\caption{The microwave-cavity response, a pump schematic, and up-converted sidebands. (a) The normalized magnitude of the signal reflected off the microwave cavity as a function of probe frequency and power. The frequency response displays a Duffing-type nonlinearity and bifurcation with increasing power. (b) The magnitude and phase as a function of frequency at a selected power in the linear regime [dashed line in (a)]. Fitting a model to these data we extract $\omega_0 / 2\pi$ = \SI{4.4403}{\giga\hertz}, total linewidth $\kappa / 2 \pi$ = \SI{1.2193}{\mega\hertz} and external linewidth $\kappa_{\text{ext}} / 2 \pi$ = \SI{0.90920}{\mega\hertz}. The two arrows mark the frequencies $\omega_\pm = \omega_0 \pm \omega_m$ of the blue- and red-detuned pumps. (c),(d) The mechanical fluctuations upconverted by the red-detuned (blue-detuned) pump, versus microwave frequency and excitation power. Solid lines show fits to a Lorentzian lineshape plus added white noise. At all pump powers, the fits reveal the same mechanical resonance frequency $\omega_m / 2 \pi$ = \SI{5.3705}{\mega\hertz} and linewidth $\gamma_m/2\pi$ = \SI{25}{\hertz}. The insets display the measured mechanical noise power versus microwave pump power converted to intracavity plasmon number $\bar{n}_c$.}%
\label{fig:measurements}
\end{figure*}

An alternative explanation of the effect of strain on superconducting thin granular films \cite{vanderLaan2007} considers the nanowire as an array of grains connected via tunneling, as depicted in Fig.~\ref{fig:device_design}(e). The Josephson inductance per tunnel junction is given in terms of its normal state tunneling resistance $L_J = R_N \hbar/\pi \Delta_0$. The two-fluid model of superconductivity \cite{Tinkham2004} leads to a similar expression for the kinetic inductance per square of film area $L_{k \square} = R_\square \hbar/\pi \Delta_0$ in terms of the normal state sheet resistance $R_\square$. Regardless of the microscopic details, substrate curvature strains the thin superconducting film, changing its tunneling resistance, sheet resistance, or energy gap, and thereby its kinetic inductance.

However, the important quantity for force transduction is the bare electromechanical coupling rate $g_0=x_{\text{zpf}}\partial\omega_0/\partial x$ of a specific device. To enhance $g_0$, the length of the nanowire should be along a region of maximum longitudinal surface strain.  A small mass $m_{\text{eff}}$ and stiffness $k$ of the mechanical mode also enhance $g_0$ by increasing the mechanical zero-point fluctuations, $x_{\text{zpf}}^2 =  \hbar/2 \sqrt{m_{\text{eff}} k}$. For AFM, the mode stiffness cannot be arbitrarily small and should be of the order of typical attractive tip-surface force gradients of approximately \SI[per-mode = symbol]{100}{\newton\per\meter}. A softer cantilever will stick to the surface at low oscillation amplitude of order \SI{1}{\nano\meter} or exhibit chaotic dynamics at large oscillation amplitudes.

Force transduction is also influenced by cavity loss rate $\kappa = \kappa_{\text{ext}} + \kappa_{\text{int}}$. The internal loss rate $\kappa_{\text{int}}$ of the cavity should be as small as possible and the cavity should be over-coupled to the external transmission line, so that cavity loss is dominated by the signal flowing to the following amplifier, $\kappa \simeq \kappa_{\text{ext}}$. The transduction efficiency is given by the dressed coupling rate $g = \sqrt{\bar{n}_c}g_0$, where $\bar{n}_c$ is the average intracavity plasmon number, which, for a given excitation power, improves with higher $Q$ of the cavity.

\section{Electromechanical coupling} 
We demonstrate KIMEC with the device shown in Fig.~\ref{fig:device_design}. A cantilever in the form of a triangular plate is formed by etching away the Si support under a Si-N layer as shown in Figs.~\ref{fig:device_design}(a) and \ref{fig:device_design}(b). Finite-element-method (FEM) simulations show the strain distribution for the fundamental flexural eigenmode is maximum along the central region of the clamping line [see Fig.~\ref{fig:device_design}(d)]. A nanowire that meanders transverse to the clamping line is etched from a thin film of niobium titanium nitride (Nb-Ti-N), forming an inductor $L_k$ that is placed in series with a coplanar capacitor $C$ to create the microwave cavity [see Fig.~\ref{fig:device_design}(a)]. An equivalent circuit is shown in Fig.~\ref{fig:device_design}(c), where all forms of internal loss at the resonance frequency $\omega_0/2\pi \simeq$ \SI{4.5}{\giga\hertz} are lumped into an effective resistance $R_0 \ll$ \SI{50}{\ohm}. On resonance, a shunt inductance $L_s \ll L_k$ transforms the 50-$\Omega$ impedance of the transmission line, bringing it closer to the impedance of the cavity $\vert Z_{\text{res}}(\omega_0) \vert = R_0$. The shunt inductance is designed to increase the loaded quality factor of the cavity while keeping it over-coupled to the transmission line, $\text{Re}[50~\Omega \parallel i \omega_0 L_s] \geq R_0$. $L_{s}$ is easily realized with a short nanowire on chip, or with a bonding wire loop between the feedline and the ground plane.

We measure the device in a dilution refrigerator cryostat with a base temperature of \SI{10}{\milli\kelvin}, well below the critical temperature $T_c \simeq$ \SI{14}{\kelvin} of Nb-Ti-N. Attenuators and isolators at various stages between room temperature and the sample stage ensure that the microwave excitation carries minimal noise. The cavity is measured in reflection using a cryogenic circulator to separate the excitation from the reflected signal, with the latter passing through a double isolator before being amplified by a cryogenic low-noise amplifier (see Appendix~\ref{sec:measurement_setup}).

Figure~\ref{fig:measurements}(a) shows the microwave response as a function of frequency and excitation power. The resonance bends toward lower frequencies and bifurcates with increasing excitation power. This nonlinear feature is caused by current-induced breaking of Cooper pairs \cite{Annunziata2010, Ku2010, Basset2019} and it is useful for parametric amplification \cite{Tholen2007, Parker2022}. At low power in the linear regime, we fit (see Appendix~\ref{sec:cavity_characterization}) the normalized reflected amplitude and phase as a function of frequency [see Fig.~\ref{fig:measurements}(b)], to extract the resonance frequency $\omega_0 / 2\pi$ = \SI{4.4403}{\giga\hertz}, total linewidth $\kappa / 2 \pi$ = \SI{1.2193}{\mega\hertz} and external linewidth $\kappa_{\text{ext}} / 2 \pi$ = \SI{0.90920}{\mega\hertz}, confirming that the cavity is over-coupled to the transmission line, as desired. The fitting gives a loaded quality factor $Q_L$ = \num{3.6417e3} and internal quality factor $Q_0$ = \num{1.4320e5}.

We probe KIMEC by exciting our device with a microwave tone, or pump, at frequency $\omega_e$, and measure the spectrum of the reflected signal. Surface strain due to cantilever motion generates sidebands in the reflected microwave signal at frequencies $\omega_e \pm \omega_m$. A pump that is red-detuned from resonance at $\omega_{-} = \omega_0 - \omega_m$ has its upper sideband at $\omega_0$ where we measure response with largest SNR, as for the lower sideband of a blue-detuned pump $\omega_{+} = \omega_0 + \omega_m$. Figures~\ref{fig:measurements}(c) and \ref{fig:measurements}(d) show the mechanical resonance detected in this manner, where we simultaneously demodulate at 72 frequencies to observe the lineshape of the mechanical mode driven by a white-noise force (see Appendix~\ref{sec:multifrequency}).

We analyze the data using a model containing two independent noise sources: the added noise of our microwave amplifier and the mechanical displacement noise of the cantilever (see Appendix~\ref{sec:noise_model}). Fitting the model to the data, we get the mechanical resonance frequency $\omega_m / 2 \pi$ = \SI{5.3705}{\mega\hertz}, in agreement with independent measurements at room temperature. At low temperatures, we obtain quality factor $Q_m$ = \num{2.1482e5} or mechanical linewidth $\gamma_m / 2 \pi$ = \SI{25}{\hertz}. Our device is in the so-called resolved-sideband regime, where $\omega_m > \kappa$. From FEM simulations, we deduce the effective mass of the fundamental flexural mode $m_{\text{eff}}$ = \SI{54}{\pico\gram}, giving a mode stiffness $k = m_{\text{eff}} \omega_m^2$ = \SI[per-mode=symbol]{61}{\newton\per\meter}. The fits confirm that the cantilever is driven by a white-noise force, i.e., frequency independent in this narrowband. However, the power spectral density of this noise force is much larger than the one expected from thermal equilibrium fluctuations at the cryostat base temperature $T_{b}$ = \SI{10}{\milli\kelvin}. Indeed, we observe no change in the magnitude of the mechanical noise up to \SI{1}{\kelvin}, above which it is difficult to stably operate the dilution refrigerator.

\begin{figure*}[htb]
\includegraphics[width=17.2cm]{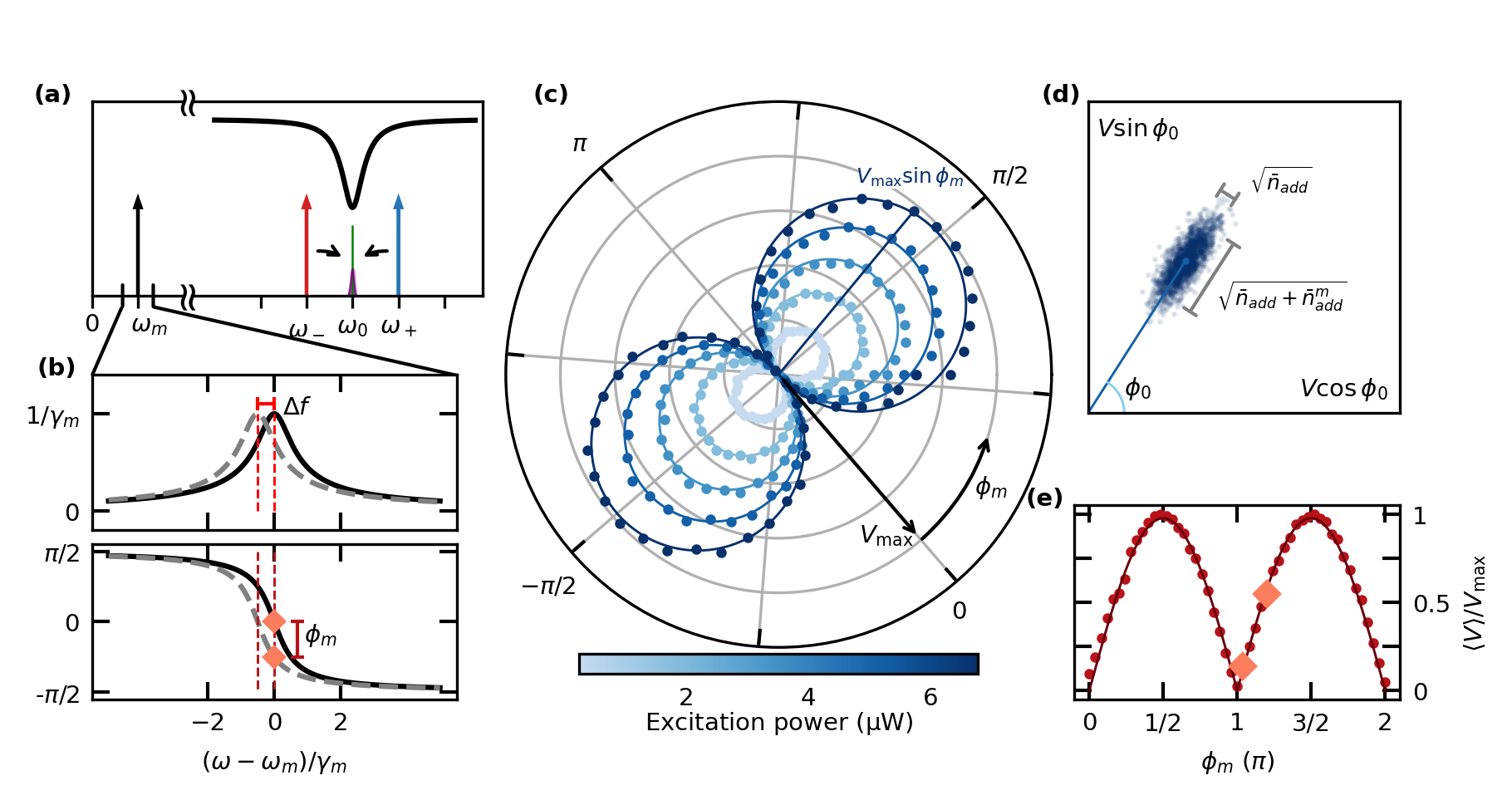}%
\caption{Phase-sensitive electromechanical transduction. (a) The multifrequency scheme for phase-sensitive detection of cantilever motion. The cantilever is driven on resonance $\omega_d \simeq \omega_m$ while simultaneously pumping the microwave cavity with two tones of equal amplitude, blue and red detuned by $\omega_d$ from the cavity resonance: $\omega_\pm = \omega_0 \pm \omega_d$. The sidebands from each pump interfere at $\omega_0$, giving the measured response. (b) A schematic showing the shift of the mechanical resonance $\Delta f$ or mechanical phase $\phi_m$ at a fixed drive frequency, due to a tip-surface force gradient. (c) A polar plot of the measured response as a function of $\phi_m$ relative to the phase $\phi_0$ of the beating wave form produced by the two pumps. (d) A scatter plot of the measured quadratures, showing the noise distribution in the phase space of the microwave cavity. Scale bars show the standard deviations of the distribution expressed as number of added noise photons $\bar{n}_{\text{add}}$ from the following amplifier and $\bar{n}_{\text{add}}^{m}$ from mechanical fluctuations. (e) The fringes resulting from the interference of the two sidebands generated by the cantilever motion and each pump. The slope of this curve at a given phase determines the responsivity of the transducer to a small shift of mechanical phase. The diamonds mark the phase shift shown in panel (b).}%
\label{fig:bae}
\end{figure*}

We can rule out heating and other sources of backaction from the microwave pump as the source of the measured mechanical noise, because the noise power depends linearly on the microwave pump power, as shown in the insets of Figs.~\ref{fig:measurements}(c) and \ref{fig:measurements}(d). Backaction would generate a mechanical noise power with a stronger-than-linear (heating) or weaker-than-linear (cooling) dependence on intracavity plasmon number. Similar devices are known to be out of equilibrium with the thermal bath at millikelvin temperatures \cite{Regal2008, Arnold2020}, but in our case it is likely that the source of this mechanical noise is the piezo actuator used to inertially actuate the cantilever. In subsequent experiments without the piezo actuator, we were not able to resolve the residual mechanical noise peak above the added background noise of our microwave amplifier. Very small noise forces are detected due to the high quality factor of the mechanical mode but we unfortunately do not know the magnitude of this residual noise force and we therefore cannot calibrate $g_0$.

Assuming an effective mode temperature $T_m$, we can estimate an upper bound on the bare electromechanical coupling rate $g_0 = x_{\text{zpf}} \partial \omega_0 / \partial x$, the associated cooperativity $C_0 = 4 g_0^2 / \kappa \gamma_m$ and the force sensitivity defined as the minimum detectable force $F_{\text{min}}$, where SNR = 1 (see Appendixes~\ref{sec:noise_model} and \ref{sec:force_sensitivity}). For $T_m$ = \SI{1}{\kelvin}, we obtain $g_0 / 2 \pi$ = \SI{9.8}{\hertz}, $C_0$ = \num{1.26e-5} and $F_{\text{min}}$ = \SI[per-mode=symbol, power-half-as-sqrt]{2.2e-17}{\newton\per\hertz\tothe{0.5}}. For $T_m$ = \SI{10}{\kelvin}, we find $g_0/2\pi$ = \SI{3.1}{\hertz}, $C_0$ = \num{1.18e-6} and $F_{\text{min}}$ = \SI[per-mode=symbol, power-half-as-sqrt]{6.8e-17}{\newton\per\hertz\tothe{0.5}}. While this coupling rate is rather small, it is nonetheless sufficient to detect the residual mechanical noise with a relatively weak microwave excitation. Larger coupling rates have been achieved with cavity optomechanical readout of cantilevers \cite{Rugar2004, Srinivasan2011} but the mode stiffness was too low for high-resolution imaging with typical tip-surface force gradients.

\section{Phase-sensitive force transduction}
Like optical cavities at room temperature, microwave cavities at millikelvin temperatures can be driven with a coherent state where the readout noise is limited by quantum fluctuations of the electromagnetic field.  Phase-sensitive detection of the mechanical oscillation (i.e., position sensing) is usually achieved by interferometry at optical frequencies. Here, we demonstrate interferometry at microwave frequencies with a multifrequency pumping scheme shown in Fig.~\ref{fig:bae}(a) which is based on backation evasion \cite{Braginsky1980, Hertzberg2010, Suh2013, Suh2014}. We apply two pumps of equal intensity, blue and red detuned at the frequencies $\omega_\pm = \omega_0 \pm \omega_d$, while measuring the response at $\omega_0$. The two pumps create a beating excitation wave form described by a carrier at $\omega_0$ which is amplitude modulated at frequency $\omega_d$, i.e., $v_e(t)=2V_{0e}\cos(\omega_d t)\cos(\omega_0t+\phi_0)$. In addition, we inertially actuate the cantilever very close to resonance, at $\omega_d \simeq \omega_m$. The cantilever motion $x(t)=\bar{X}\sin(\omega_d t +\phi_m)$ generates an upper sideband of the red pump and a lower sideband of the blue pump, which interfere at the cavity resonance frequency $\omega_0$ [see Fig.~\ref{fig:bae}(a)]. The device is effectively operating as an interferometer, generating a microwave response proportional to the amplitude of a single quadrature of the mechanical oscillation. Figure~\ref{fig:bae}(c) shows a polar plot of the averaged signal amplitude as a function of the mechanical phase for different pump powers. The interference of the sidebands produced by the two pumps gives rise to the dependence on mechanical phase, displayed for a single pump power as fringes in Fig.~\ref{fig:bae}(e). To perform this measurement, we use a multifrequency microwave lock-in amplifier \cite{imp} and tune the measurement frequencies $\omega_d$, $\omega_\pm$, and $\omega_0$, to establish one common phase reference, with independent control of the carrier phase $\phi_0$ and the mechanical phase $\phi_m$ (see Appendix~\ref{sec:multifrequency}).

The measurement gives both quadratures of the microwave response, which depend on the mechanical phase, $V(\omega_0)=V_{\text{max}}\sin (\phi_m)+ V_{\text{noise}}$. Figure~\ref{fig:bae}(d) shows a scatter plot of 2000 measurements with bandwidth $\simeq \gamma_m$. The fluctuations of the quadratures $V_{\text{noise}}$ show an elliptical Gaussian distribution, the result of phase-sensitive detection, where the major axis of the noise ellipse is dominated by mechanical noise and the minor axis by the added noise of the following microwave amplifier.

When operating as a force sensor, the cantilever is sensitive to changes in the tip-surface force gradient $F'$. The tip-surface interaction modifies the effective stiffness of the mechanical mode, shifting the mechanical resonance frequency [see Fig.~\ref{fig:bae}(b)]. To first order, the shift in $\omega_m$, or change in $\phi_m$, is proportional to the mechanical quality factor $Q_m$. The two-tone excitation maps this mechanical phase shift to a change in the average value $\langle V \rangle $, with a responsivity that depends on the excitation power $P_{\text{in}}$ and the bare coupling rate $g_0$. The force gradient responsivity is given by (see Appendix~\ref{sec:force_sensitivity})
\begin{align}
\begin{split}
    \frac{\partial\langle V \rangle}{\partial F'} \bigg\rvert_{\phi_m=0} &= \sqrt{P_{\text{in}} Z_{\text{ext}}} \frac{4 \kappa_{\text{ext}}}{\sqrt{\kappa^2 (\kappa^2 + 4 \omega_m^2)}} g_0 \frac{\bar{X}}{x_{\text{zpf}}} \\ 
    & \times \frac{\omega_m}{\gamma_m(k+F')},
    \label{eq:force_gradient_responsivity}
\end{split}
\end{align}
where $Z_{\text{ext}}$ is the 50 $\Omega$ impedance of the source and $\bar{X}$ the amplitude of the mechanical oscillation. Thus, the constant $g_0$ and the cantilever stiffness $k$ are the calibration constants that connect an electrical measurement in the microwave frequency band to a measurement of force on the AFM cantilever.

\section{Conclusion}
We have demonstrated a force transducer based on a type of electromechanical coupling where strain modifies the kinetic inductance of a superconducting nanowire. Our device was designed for application in low-temperature atomic force microscopy but the compact coplanar design and ease of fabrication makes it attractive for several different applications. We realized a device in the resolved-sideband regime and we have shown how to implement phase-sensitive detection of the cantilever oscillation  with a multifrequency drive and detection scheme. Our measurements show the presence of a residual white-noise force that is not in equilibrium with the cryostat at base temperature. We estimate a bare coupling rate $g_0/2\pi$ in the range \SIrange{3}{10}{\hertz} corresponding to a single-plasmon cooperativity $C_0$ in the range \numrange{1.18e-6}{1.26e-5}. Our ability to reach stronger dressed coupling $g = \sqrt{\bar{n}_c} g_0$ is limited by the bifurcation of the cavity at intracavity plasmon number $\bar n_c \sim 10^5$, due to current-dependent Cooper-pair breaking. Such nonlinearity may be useful in future investigations for the integration of microwave parametric amplification to increase detection sensitivity.

\section*{Acknowledgments}
We would like to thank the Quantum-Limited Atomic Force Microscopy (QAFM) team, T. Glatzel, M. Zutter, E. Tholén, D. Forchheimer, I. Ignat, M. Kwon, J. Hafner and D. Platz, for fruitful discussions. We would additionally like to acknowledge Joe Aumentado and John Teufel at National Institute of Standards and Technology (NIST) Boulder for helpful discussions. This work was supported by an EIC Pathfinder Grant No. 828966---QAFM and the Swedish SSF Grant No. ITM17-0343.

\section*{Data availability}
The data that support the findings of this study and the code used to generate the figures for this manuscript are available on Zenodo \cite{data_repo}.

\section*{Author contributions}
A.K.R., E.S. and E.K.A. performed the measurements, analyzed the data and wrote the manuscript with E.H. and D.B.H. A.K.R., E.S. and E.K.A. designed and fabricated the devices with E.H. D.B.H. conceived the experiment and supervised the project.

\section*{Competing interests}
D.B.H. is a part owner of Intermodulation Products AB, which manufactures and sells the microwave measurement platform used in this experiment. A.K.R, E.S., E.K.A., E.H. declare no competing interests.

\begin{appendix}

\section{Device fabrication}\label{sec:fabrication}
We fabricate our devices on 100-mm-diameter, 525-$\mu$m-thick silicon (Si) wafers with a coating of 600-nm-thick ($\pm 5\%$) super-low-stress silicon nitride (Si-N). On the front side, we sputter a 15-nm-thick film of niobium titanium nitride (Nb-Ti-N). Using photolithography, we pattern metal contact pads, alignment markers, and, on the back side of the wafer, a metal mask used for the final deep silicon etch. We use electron-beam-lithography and photolithography, together with dry etching, to pattern the inductor and capacitor, as well as the rest of the electrical circuit. To shape each chip and define the Si-N cantilever, we use photolithography and dry etching on the front side. A deep silicon etch from the back side separates each chip from the wafer, and a final KOH silicon etch from the front side releases the Si-N cantilevers.

\section{Mechanical and Electromagnetic Modeling}\label{sec:modeling}
We model the fundamental flexural mode of the triangular cantilever using finite-element-methods (FEM) with COMSOL Multiphysics \cite{comsol}. The dimensions of the cantilever model are fixed to the fabricated values. The model gives the normalized volumetric strain profile of the cantilever and effective mass $m_{\text{eff}}$, assuming the bulk density $\rho_{\text{Si-N}}$ = \SI[per-mode=symbol]{3100}{\kilo\gram\per\meter\cubed}.

We model the electromagnetic resonances with SONNET, which implements a quasi-three-dimensional electromagnetic model that accounts for sheet kinetic inductance \cite{sonnet}. We find good correspondence between the model and the measured resonance frequencies with $L_{k\square}$ = \SI[per-mode=symbol]{38}{\pico\henry\per\sq}, similar to reported values for approximately 100 nm $\times$ 100-nm square size \cite{Samkharadze2016}. For a given inductor design the simulations verify that the frequency of the fundamental self-resonance lies well above the resonance frequency of the lumped-element $L_k C$ mode.

\section{Measurement setup}\label{sec:measurement_setup}
\begin{figure}[ht]
\includegraphics[width=8.6cm]{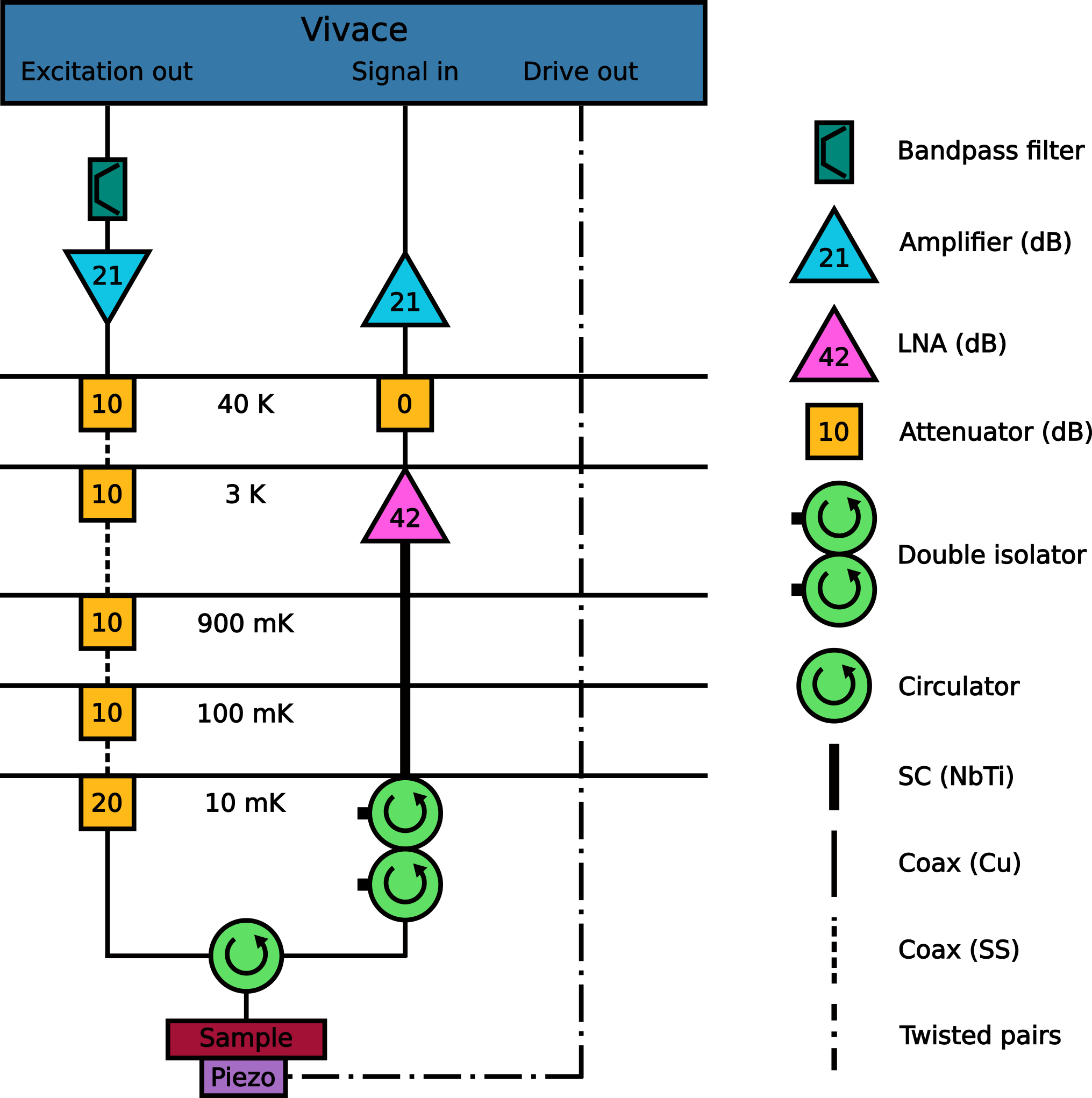}%
\caption{Schematic of the measurement setup. The materials and parameters of each component is given in the legend.}%
\label{fig:setup}
\end{figure}

Figure~\ref{fig:setup} details the measurement setup for our experiments. The sample is glued to a piezoelectric disk and wire bonded to a printed circuit board (PCB). The piezo shakes the entire chip, putting it in a noninertial reference frame to inertially actuate cantilever oscillation. The low-frequency drive (approximately \SI{5}{\mega\hertz}) to the piezo is supplied via a filtered twisted pair. The entire device is thermally anchored to the mixing-chamber plate of a dilution refrigerator (DR) and cooled down to \SI{10}{\milli\kelvin}. The superconducting resonant circuit is measured in reflection with a cryogenic circulator at the 10-mK stage to separate the incoming and reflected wave. A double isolator after the circulator protects the cavity from noise injected by the following microwave amplifiers. The added noise of the amplification chain is determined by the first amplifier, a cryogenic low noise amplifier (LNF-LNA4\textunderscore8C, 42-dB gain), mounted on the 3-K plate. The readout line is semirigid copper coaxial cable at the 10-mK stage and semirigid Nb-Ti superconducting coaxial cables between subsequent temperature stages, to minimize thermal conduction and signal loss. To reduce thermal conduction, the microwave pump line is a stainless-steel semirigid coaxial cable between different temperature stages, with cold attenuators at each stage to thermalize the noise from higher temperature.

\section{Cavity characterization}\label{sec:cavity_characterization}
The ideal response of a single mode of a microwave cavity directly coupled to a transmission line and measured in reflection is given by Ref.~\cite{Pozar2011} as
\begin{equation}
    S_{11}(\omega) = 1 - \frac{\kappa_{\text{ext}}}{\frac{\kappa}{2} - i (\omega_0 - \omega)},
    \label{eq:S11}
\end{equation}
where $\omega_0$ is the resonance frequency, $\kappa_{\text{ext}}$ is the external linewidth and $\kappa$ is the total linewidth. External microwave components and cables introduce background losses and distortions to this ideal frequency response. Similar to the methods described in Refs.~\cite{Fink2016, Bothner2020}, we model the combination of the ideal cavity and the background with
\begin{equation}
  S_{11}(\omega) = (\alpha_0 + \alpha_1 \omega) \left(1 - \frac{\kappa_{\text{ext}}e^{i\theta}}{\frac{\kappa}{2} - i (\omega_0 - \omega)} \right) e^{i(\beta_0 + \beta_1 \omega)},
  \label{eqn:modified_cavity_response}
\end{equation}
where $\alpha_0, \alpha_1, \beta_0, \beta_1$ account for an amplitude offset and linear frequency-dependent attenuation and a phase offset and linear frequency-dependent phase shift (i.e., electrical delay). The factor $e^{i \theta}$ captures the asymmetry of the linewidth for low probe powers in the linear regime of the cavity. Using a least-squares minimization method \cite{lmfit}, we fit Eq.~\eqref{eqn:modified_cavity_response} to the measured frequency response of the cavity, extracting $\omega_0, \kappa$ and $\kappa_{\text{ext}}$ given in the main text.

\section{Multifrequency techniques}\label{sec:multifrequency}
All measurement signals were synthesized and measured with the Vivace microwave measurement platform from Intermodulation Products AB \cite{imp}, running firmware that implements a multifrequency lock-in amplifier with up to 144 frequency components (tones) distributed over eight output and eight input ports. The tones are set either in the baseband (\SIrange{0}{500}{\mega\hertz}) or up- or down-converted to the signal band with a radio-frequency numerically controlled oscillator (NCO) via digital I-Q mixing. We tune all tones of interest to be integer multiples of the measurement bandwidth $\Delta f$, i.e., the inverse of the measurement time-window used to calculate Fourier sums. This tuning eliminates Fourier leakage between closely spaced tones and provides a common phase reference for all frequencies, including intermodulation or frequency-mixing products. For the data shown in Fig.~\ref{fig:measurements}(c) and \ref{fig:measurements}(d), at each excitation power we simultaneously measure both quadratures at 72 tones to reveal the frequency-dependent magnitude and random phase produced by mechanical fluctuations, with each tone separated by the measurement bandwidth $\Delta f$ = \SI{8}{\hertz} $<\gamma_m / 2\pi$.

For the measurement in Fig.~\ref{fig:bae}(a), Vivace generates the two-tone microwave excitation with frequency components $\omega_\pm = \omega_0 \pm \omega_d$ on one port, each with independent phase control $\phi_+$ and $\phi_-$. The mechanical drive signal at $\omega_d$ is generated on another port, also with independent phase control $\phi_{md}$. The common-mode phase of the microwave excitation is $\phi_0 = \frac{1}{2} (\phi_+ +\phi_-)$ and the differential phase $\phi_m^e = \frac{1}{2} (\phi_+-\phi_-)$, where the electrical delay only affects $\phi_0$. $\phi_m^e$ is the phase of the beating microwave wave form that establishes the reference for measurement of the phase of the mechanical response $\phi_m = \phi_{md} + \Delta \phi - \phi_m^e$, where $\Delta \phi$ is a constant phase offset that depends on the electrical delay and the mechanical susceptibility.

For the data in Figs.~\ref{fig:bae}(c)--(e), we measure a single lock-in component at the microwave-cavity resonance frequency while stepping the applied phase of the mechanical drive $\phi_{md}$. At each drive phase, we acquire 2000 I and Q quadratures of the microwave response with an integration bandwidth $\Delta f$ = \SI{25}{\hertz} $\simeq \gamma_m/2\pi$ to obtain the noise distribution in phase space.

\section{Noise model}\label{sec:noise_model}
When the cavity is excited at a detuned frequency $ \omega_e = \omega_0 + \Delta $, the power spectral density of mechanical displacement noise at the frequency $\Omega$, $S_{xx}(\Omega)$, is related to fluctuations in the cavity resonance frequency $S_{\omega \omega}(\Omega)$:
\begin{equation}
    S_{\omega \omega}(\Omega) = \frac{g_0^{2} \bar n_c(\Delta)}{x_{\text{zpf}}^{2}} S_{xx}(\Omega).
\label{eq:Sww}
\end{equation}

Here, $\bar n_c(\Delta)$ is the number of intracavity plasmons, determined by the input power $P_{\text{in}}$ and the detuning $\Delta$,
\begin{equation}
    \bar n_c(\Delta) = \frac{P_{\text{in}}}{\hbar \omega_e} \vert \mathcal{K}(\Delta) \vert^{2},
\label{eq:nc}
\end{equation}
where $\mathcal{K}(\delta)$ is the cavity susceptibility at a frequency $\delta$ offset from resonance,
\begin{equation}
    \mathcal{K}(\delta) = \frac{\sqrt{\kappa_{\text{ext}}}}{\frac{\kappa}{2} + i \delta}.
\label{eq:K}
\end{equation}

Fluctuations of the cavity resonance frequency generate voltage fluctuations at the device port, in two frequency sidebands centered at the excitation frequency,
\begin{equation}
    S_{VV}(\omega_e \pm \Omega)  = \hbar \omega_e Z_{\text{ext}} \vert \mathcal{K}(\Delta \pm \Omega) \vert^{2} S_{\omega \omega}(\Omega),
    \label{eq:SVV2}
\end{equation}
where $Z_{\text{ext}}$ is the impedance loading the device port. Combining Eqs.~\eqref{eq:Sww} and \eqref{eq:SVV2}, we relate the voltage noise power spectral density to the power spectral density of the mechanical displacement noise:
\begin{equation}
     S_{VV}(\omega_e \pm \Omega) =  P_{\text{in}} Z_{\text{ext}} \vert \mathcal{K}(\Delta \pm \Omega) \vert^{2}  \vert \mathcal{K}(\Delta) \vert^{2}  \frac{g_0^{2}}{x_{\text{zpf}}^{2}} S_{xx}(\Omega).
\end{equation}

When measuring far from resonance, the input power is completely reflected and $P_{\text{in}} = P_{r}$. A red-detuned excitation $\Delta = - \omega_m$ and narrow-band probe $\Omega \simeq +\omega_m$ at the upper sideband, on the cavity resonance, gives \cite{Golokolenov2021}
\begin{equation}
     S_{VV}(\omega_e + \Omega) = P_r Z_{\text{ext}}   \frac{16 \kappa_{\text{ext}}^{2}}{\kappa^2 (\kappa^2 + 4 \omega_m^2)} \frac{g_0^{2}}{x_{\text{zpf}}^{2}}S_{xx}(\Omega),
\end{equation}
which relates the microwave voltage noise spectrum $S_{VV}$ to the low-frequency mechanical fluctuations $S_{xx}$.

For a single mechanical eigenmode, or simple harmonic oscillator, in thermal equilibrium with bath at temperature $T_m$, the fluctuation-dissipation theorem gives,
\begin{equation}
    \frac{S_{xx}(\Omega)}{x_{\text{zpf}}^2}= \frac{k_BT_m}{\hbar \omega_m}\frac{\gamma_m}{(\Omega-\omega_m)^2+\gamma_m^2/4}.
\end{equation}
We measure the amplified power spectral density of microwave voltage fluctuations at the top of the cryostat,
\begin{equation}
     S_{VV}^\text{measured}(\omega) = G \left[ S_{VV}(\omega) + S_{VV}^\text{add} \right],
\end{equation}
where $G$ is the gain and $S_{VV}^\text{add}$ is the added noise of our amplification chain. Therefore, we relate the measured microwave voltage noise spectrum to our device parameters and environment as
\begin{equation}
\begin{split}
    \frac{S_{VV}^{\text{measured}(\omega)}}{P_{\text{measured}}} &= Z_{\text{ext}} \frac{16 \kappa_{\text{ext}}^{2}}{\kappa^2 (\kappa^2 + 4 \omega_m^2)} 
    \frac{g_0^{2} k_BT_m}{\hbar \omega_m}\\
    & \times \frac{\gamma_m}{(\Omega-\omega_m)^2+\gamma_m^2/4}
    + \text{const.},
\label{eq:SVVmeasured}
\end{split}
\end{equation}
where $P_{\text{measured}} = G P_r$ is the measured reflected power and the added noise is constant, or independent of frequency. We measure the left-hand side of Eq.~\eqref{eq:SVVmeasured} and fit the frequency dependence expressed in the right-hand side, to determine $g_0$. All parameters except for $T_m$ on the right-hand side are independently determined.

When the motion is coherently driven at the mechanical resonance frequency, the response in the phase-sensitive detection scheme is expressed as
\begin{equation}
    \langle V \rangle = \sqrt{P_{\text{in}}Z_{\text{ext}}}\frac{4\kappa_{\text{ext}}}{\sqrt{\kappa^2(\kappa^2+4\omega_d^2)}} g_0 \frac{\hat{X}\delta(\Omega - \omega_d)}{x_{\text{zpf}}}\sin{\phi_m},
\label{eq:phase_sensitive_detection}
\end{equation}
where $\delta(\omega)$ is the Dirac delta function.

\section{Sensitivity of force and position measurement}\label{sec:force_sensitivity}
The force sensitivity is expressed as the minimum detectable force, or force detected at SNR = 1. The limitations imposed by the onset of the nonlinearity of the cavity set the upper bound for the pump power and at this upper bound the effect of backation can be neglected. We therefore neglect backation noise and consider only two contributions: mechanical fluctuations at an effective temperature $T_m$ and the added detector noise. The minimum detectable force is given by \cite{Giessibl2003, Hertzberg2010},
\begin{equation}
    F_{\text{min}} = m_{\text{eff}} \omega_m \gamma_m \sqrt{S_{xx}^{\text{add}} + \frac{4 k_{B} T_m}{m_{\text{eff}} \omega_m^2 \gamma_m}},
\end{equation}
where the added noise $S_{xx}^{\text{add}}$ is expressed as an equivalent deflection noise in the bandwidth $\text{B}$,
\begin{equation}
     S_{xx}^{\text{add}} = \frac{\text{B}}{\hbar \omega Z_{\text{ext}} \bar{n}_{\text{add}}} \frac{x_{\text{zpf}}^2}{g_0^2} S_{VV}^{\text{add}}.
\end{equation}
The attenuation chain ensures that the quantum fluctuations of the electromagnetic field at microwave frequency $\omega$ are close to ground state as $k_B T_{b}\ll \hbar \omega_0$ at $T_b$ = \SI{10}{\milli\kelvin}. In our setup, the added noise from the following cryogenic low-noise amplifier is the dominant contribution. Our amplifier (LNF-LNC4\textunderscore8C) has noise temperature $T_{N} \approx$ \SI{2.5}{\kelvin} corresponding to an average of $\bar{n}_{\text{add}} = k_{B} T_{N} / \hbar\omega_0 \approx 12$~photons of added noise. For reasonably large pump powers, the spectral density due to the thermal mechanical motion of the cantilever dominates the background noise (see Fig.~\ref{fig:measurements}) so that $S_{xx}^{\text{add}}$ can be neglected and the force sensitivity reduces to
\begin{equation}
    F_{\text{min}} = \sqrt{4 k_{B} T_m m_{\text{eff}} \gamma_m},
\end{equation}
which, using the measured values of $\gamma_m/2\pi$ = \SI{25}{\hertz} and $m_{\text{eff}}$ = \SI{54}{\pico\gram}, yields $F_{\text{min}}$ = \SI[per-mode=symbol, power-half-as-sqrt]{2.2e-17}{\newton\per\hertz\tothe{0.5}}, assuming $T_m$ = \SI{1}{K}.

The force gradient responsivity expressed in Eq.~\eqref{eq:force_gradient_responsivity} is derived as the product of three contributions:
\begin{align}
    \frac{\partial \langle V \rangle}{\partial F'}=\frac{\partial\langle V \rangle}{\partial \phi _m}\frac{\partial \phi_m}{\partial \omega_m}\frac{\partial \omega_m}{\partial F'}.
\end{align}
The first term follows from Eq.~\eqref{eq:phase_sensitive_detection}, while the second term stems from the change of the mechanical phase due to a frequency shift for a fixed drive frequency, expressed as the slope of the phase of the mechanical susceptibility on resonance $\partial \phi_m/\partial \omega = 2 /\gamma_m$. The third term comes from the change of the effective mechanical resonance frequency $\omega_m=\sqrt{(k+F')/m}$ due to a force gradient $\partial \omega_m/\partial F'=\omega_m/2(k+F')$.

\end{appendix}

\providecommand{\noopsort}[1]{}\providecommand{\singleletter}[1]{#1}%


\begin{thebibliography}{43}%
\makeatletter
\providecommand \@ifxundefined [1]{%
 \@ifx{#1\undefined}
}%
\providecommand \@ifnum [1]{%
 \ifnum #1\expandafter \@firstoftwo
 \else \expandafter \@secondoftwo
 \fi
}%
\providecommand \@ifx [1]{%
 \ifx #1\expandafter \@firstoftwo
 \else \expandafter \@secondoftwo
 \fi
}%
\providecommand \natexlab [1]{#1}%
\providecommand \enquote  [1]{``#1''}%
\providecommand \bibnamefont  [1]{#1}%
\providecommand \bibfnamefont [1]{#1}%
\providecommand \citenamefont [1]{#1}%
\providecommand \href@noop [0]{\@secondoftwo}%
\providecommand \href [0]{\begingroup \@sanitize@url \@href}%
\providecommand \@href[1]{\@@startlink{#1}\@@href}%
\providecommand \@@href[1]{\endgroup#1\@@endlink}%
\providecommand \@sanitize@url [0]{\catcode `\\12\catcode `\$12\catcode
  `\&12\catcode `\#12\catcode `\^12\catcode `\_12\catcode `\%12\relax}%
\providecommand \@@startlink[1]{}%
\providecommand \@@endlink[0]{}%
\providecommand \url  [0]{\begingroup\@sanitize@url \@url }%
\providecommand \@url [1]{\endgroup\@href {#1}{\urlprefix }}%
\providecommand \urlprefix  [0]{URL }%
\providecommand \Eprint [0]{\href }%
\providecommand \doibase [0]{https://doi.org/}%
\providecommand \selectlanguage [0]{\@gobble}%
\providecommand \bibinfo  [0]{\@secondoftwo}%
\providecommand \bibfield  [0]{\@secondoftwo}%
\providecommand \translation [1]{[#1]}%
\providecommand \BibitemOpen [0]{}%
\providecommand \bibitemStop [0]{}%
\providecommand \bibitemNoStop [0]{.\EOS\space}%
\providecommand \EOS [0]{\spacefactor3000\relax}%
\providecommand \BibitemShut  [1]{\csname bibitem#1\endcsname}%
\let\auto@bib@innerbib\@empty
\bibitem [{\citenamefont {Binnig}\ \emph {et~al.}(1986)\citenamefont {Binnig},
  \citenamefont {Quate},\ and\ \citenamefont {Gerber}}]{Binning1986}%
  \BibitemOpen
  \bibfield  {author} {\bibinfo {author} {\bibfnamefont {G.}~\bibnamefont
  {Binnig}}, \bibinfo {author} {\bibfnamefont {C.~F.}\ \bibnamefont {Quate}},\
  and\ \bibinfo {author} {\bibfnamefont {C.}~\bibnamefont {Gerber}},\
  }\bibfield  {title} {\bibinfo {title} {Atomic force microscope},\ }\href
  {https://doi.org/10.1103/PhysRevLett.56.930} {\bibfield  {journal} {\bibinfo
  {journal} {Phys. Rev. Lett.}\ }\textbf {\bibinfo {volume} {56}},\ \bibinfo
  {pages} {930} (\bibinfo {year} {1986})}\BibitemShut {NoStop}%
\bibitem [{\citenamefont {Giessibl}(2003)}]{Giessibl2003}%
  \BibitemOpen
  \bibfield  {author} {\bibinfo {author} {\bibfnamefont {F.~J.}\ \bibnamefont
  {Giessibl}},\ }\bibfield  {title} {\bibinfo {title} {Advances in atomic force
  microscopy},\ }\href {https://doi.org/10.1103/RevModPhys.75.949} {\bibfield
  {journal} {\bibinfo  {journal} {Rev. Mod. Phys.}\ }\textbf {\bibinfo {volume}
  {75}},\ \bibinfo {pages} {949} (\bibinfo {year} {2003})}\BibitemShut
  {NoStop}%
\bibitem [{\citenamefont {Braginskiĭ}\ and\ \citenamefont
  {Vorontsov}(1975)}]{Braginsky1975}%
  \BibitemOpen
  \bibfield  {author} {\bibinfo {author} {\bibfnamefont {V.~B.}\ \bibnamefont
  {Braginskiĭ}}\ and\ \bibinfo {author} {\bibfnamefont {Y.~I.}\ \bibnamefont
  {Vorontsov}},\ }\bibfield  {title} {\bibinfo {title} {Quantum-mechanical
  limitations in macroscopic experiments and modern experimental technique},\
  }\href {https://doi.org/10.1070/PU1975v017n05ABEH004362} {\bibfield
  {journal} {\bibinfo  {journal} {Soviet Physics Uspekhi}\ }\textbf {\bibinfo
  {volume} {17}},\ \bibinfo {pages} {644} (\bibinfo {year} {1975})}\BibitemShut
  {NoStop}%
\bibitem [{\citenamefont {Aspelmeyer}\ \emph {et~al.}(2014)\citenamefont
  {Aspelmeyer}, \citenamefont {Kippenberg},\ and\ \citenamefont
  {Marquardt}}]{Aspelmeyer2014}%
  \BibitemOpen
  \bibfield  {author} {\bibinfo {author} {\bibfnamefont {M.}~\bibnamefont
  {Aspelmeyer}}, \bibinfo {author} {\bibfnamefont {T.~J.}\ \bibnamefont
  {Kippenberg}},\ and\ \bibinfo {author} {\bibfnamefont {F.}~\bibnamefont
  {Marquardt}},\ }\bibfield  {title} {\bibinfo {title} {Cavity optomechanics},\
  }\href {https://doi.org/10.1103/RevModPhys.86.1391} {\bibfield  {journal}
  {\bibinfo  {journal} {Rev. Mod. Phys.}\ }\textbf {\bibinfo {volume} {86}},\
  \bibinfo {pages} {1391} (\bibinfo {year} {2014})}\BibitemShut {NoStop}%
\bibitem [{\citenamefont {Rugar}\ and\ \citenamefont
  {Gr\"utter}(1991)}]{Rugar1991}%
  \BibitemOpen
  \bibfield  {author} {\bibinfo {author} {\bibfnamefont {D.}~\bibnamefont
  {Rugar}}\ and\ \bibinfo {author} {\bibfnamefont {P.}~\bibnamefont
  {Gr\"utter}},\ }\bibfield  {title} {\bibinfo {title} {Mechanical parametric
  amplification and thermomechanical noise squeezing},\ }\href
  {https://doi.org/10.1103/PhysRevLett.67.699} {\bibfield  {journal} {\bibinfo
  {journal} {Phys. Rev. Lett.}\ }\textbf {\bibinfo {volume} {67}},\ \bibinfo
  {pages} {699} (\bibinfo {year} {1991})}\BibitemShut {NoStop}%
\bibitem [{\citenamefont {Smith}(1995)}]{Smith1995}%
  \BibitemOpen
  \bibfield  {author} {\bibinfo {author} {\bibfnamefont {D.~P.~E.}\
  \bibnamefont {Smith}},\ }\bibfield  {title} {\bibinfo {title} {Limits of
  force microscopy},\ }\href {https://doi.org/10.1063/1.1145550} {\bibfield
  {journal} {\bibinfo  {journal} {Review of Scientific Instruments}\ }\textbf
  {\bibinfo {volume} {66}},\ \bibinfo {pages} {3191} (\bibinfo {year}
  {1995})},\ \Eprint {https://arxiv.org/abs/https://doi.org/10.1063/1.1145550}
  {https://doi.org/10.1063/1.1145550} \BibitemShut {NoStop}%
\bibitem [{\citenamefont {Srinivasan}\ \emph {et~al.}(2011)\citenamefont
  {Srinivasan}, \citenamefont {Miao}, \citenamefont {Rakher}, \citenamefont
  {Davanço},\ and\ \citenamefont {Aksyuk}}]{Srinivasan2011}%
  \BibitemOpen
  \bibfield  {author} {\bibinfo {author} {\bibfnamefont {K.}~\bibnamefont
  {Srinivasan}}, \bibinfo {author} {\bibfnamefont {H.}~\bibnamefont {Miao}},
  \bibinfo {author} {\bibfnamefont {M.~T.}\ \bibnamefont {Rakher}}, \bibinfo
  {author} {\bibfnamefont {M.}~\bibnamefont {Davanço}},\ and\ \bibinfo
  {author} {\bibfnamefont {V.}~\bibnamefont {Aksyuk}},\ }\bibfield  {title}
  {\bibinfo {title} {Optomechanical transduction of an integrated silicon
  cantilever probe using a microdisk resonator},\ }\href
  {https://doi.org/10.1021/nl104018r} {\bibfield  {journal} {\bibinfo
  {journal} {Nano Letters}\ }\textbf {\bibinfo {volume} {11}},\ \bibinfo
  {pages} {791} (\bibinfo {year} {2011})},\ \bibinfo {note} {pMID: 21250747},\
  \Eprint {https://arxiv.org/abs/https://doi.org/10.1021/nl104018r}
  {https://doi.org/10.1021/nl104018r} \BibitemShut {NoStop}%
\bibitem [{\citenamefont {H\"alg}\ \emph {et~al.}(2021)\citenamefont {H\"alg},
  \citenamefont {Gisler}, \citenamefont {Tsaturyan}, \citenamefont {Catalini},
  \citenamefont {Grob}, \citenamefont {Krass}, \citenamefont {H\'eritier},
  \citenamefont {Mattiat}, \citenamefont {Thamm}, \citenamefont {Schirhagl},
  \citenamefont {Langman}, \citenamefont {Schliesser}, \citenamefont {Degen},\
  and\ \citenamefont {Eichler}}]{Halg2021}%
  \BibitemOpen
  \bibfield  {author} {\bibinfo {author} {\bibfnamefont {D.}~\bibnamefont
  {H\"alg}}, \bibinfo {author} {\bibfnamefont {T.}~\bibnamefont {Gisler}},
  \bibinfo {author} {\bibfnamefont {Y.}~\bibnamefont {Tsaturyan}}, \bibinfo
  {author} {\bibfnamefont {L.}~\bibnamefont {Catalini}}, \bibinfo {author}
  {\bibfnamefont {U.}~\bibnamefont {Grob}}, \bibinfo {author} {\bibfnamefont
  {M.-D.}\ \bibnamefont {Krass}}, \bibinfo {author} {\bibfnamefont
  {M.}~\bibnamefont {H\'eritier}}, \bibinfo {author} {\bibfnamefont
  {H.}~\bibnamefont {Mattiat}}, \bibinfo {author} {\bibfnamefont {A.-K.}\
  \bibnamefont {Thamm}}, \bibinfo {author} {\bibfnamefont {R.}~\bibnamefont
  {Schirhagl}}, \bibinfo {author} {\bibfnamefont {E.~C.}\ \bibnamefont
  {Langman}}, \bibinfo {author} {\bibfnamefont {A.}~\bibnamefont {Schliesser}},
  \bibinfo {author} {\bibfnamefont {C.~L.}\ \bibnamefont {Degen}},\ and\
  \bibinfo {author} {\bibfnamefont {A.}~\bibnamefont {Eichler}},\ }\bibfield
  {title} {\bibinfo {title} {Membrane-based scanning force microscopy},\ }\href
  {https://doi.org/10.1103/PhysRevApplied.15.L021001} {\bibfield  {journal}
  {\bibinfo  {journal} {Phys. Rev. Appl.}\ }\textbf {\bibinfo {volume} {15}},\
  \bibinfo {pages} {L021001} (\bibinfo {year} {2021})}\BibitemShut {NoStop}%
\bibitem [{\citenamefont {Nation}\ \emph {et~al.}(2016)\citenamefont {Nation},
  \citenamefont {Suh},\ and\ \citenamefont {Blencowe}}]{Nation2016}%
  \BibitemOpen
  \bibfield  {author} {\bibinfo {author} {\bibfnamefont {P.~D.}\ \bibnamefont
  {Nation}}, \bibinfo {author} {\bibfnamefont {J.}~\bibnamefont {Suh}},\ and\
  \bibinfo {author} {\bibfnamefont {M.~P.}\ \bibnamefont {Blencowe}},\
  }\bibfield  {title} {\bibinfo {title} {Ultrastrong optomechanics
  incorporating the dynamical casimir effect},\ }\href
  {https://doi.org/10.1103/PhysRevA.93.022510} {\bibfield  {journal} {\bibinfo
  {journal} {Phys. Rev. A}\ }\textbf {\bibinfo {volume} {93}},\ \bibinfo
  {pages} {022510} (\bibinfo {year} {2016})}\BibitemShut {NoStop}%
\bibitem [{\citenamefont {Rodrigues}\ \emph {et~al.}(2019)\citenamefont
  {Rodrigues}, \citenamefont {Bothner},\ and\ \citenamefont
  {Steele}}]{Rodrigues2019}%
  \BibitemOpen
  \bibfield  {author} {\bibinfo {author} {\bibfnamefont {I.~C.}\ \bibnamefont
  {Rodrigues}}, \bibinfo {author} {\bibfnamefont {D.}~\bibnamefont {Bothner}},\
  and\ \bibinfo {author} {\bibfnamefont {G.~A.}\ \bibnamefont {Steele}},\
  }\bibfield  {title} {\bibinfo {title} {Coupling microwave photons to a
  mechanical resonator using quantum interference},\ }\href
  {https://doi.org/10.1038/s41467-019-12964-2} {\bibfield  {journal} {\bibinfo
  {journal} {Nature Communications}\ }\textbf {\bibinfo {volume} {10}},\
  \bibinfo {pages} {5359} (\bibinfo {year} {2019})}\BibitemShut {NoStop}%
\bibitem [{\citenamefont {Regal}\ \emph {et~al.}(2008)\citenamefont {Regal},
  \citenamefont {Teufel},\ and\ \citenamefont {Lehnert}}]{Regal2008}%
  \BibitemOpen
  \bibfield  {author} {\bibinfo {author} {\bibfnamefont {C.~A.}\ \bibnamefont
  {Regal}}, \bibinfo {author} {\bibfnamefont {J.~D.}\ \bibnamefont {Teufel}},\
  and\ \bibinfo {author} {\bibfnamefont {K.~W.}\ \bibnamefont {Lehnert}},\
  }\bibfield  {title} {\bibinfo {title} {Measuring nanomechanical motion with a
  microwave cavity interferometer},\ }\href {https://doi.org/10.1038/nphys974}
  {\bibfield  {journal} {\bibinfo  {journal} {Nature Physics}\ }\textbf
  {\bibinfo {volume} {4}},\ \bibinfo {pages} {555} (\bibinfo {year}
  {2008})}\BibitemShut {NoStop}%
\bibitem [{\citenamefont {Teufel}\ \emph {et~al.}(2009)\citenamefont {Teufel},
  \citenamefont {Donner}, \citenamefont {Castellanos-Beltran}, \citenamefont
  {Harlow},\ and\ \citenamefont {Lehnert}}]{Teufel2009}%
  \BibitemOpen
  \bibfield  {author} {\bibinfo {author} {\bibfnamefont {J.~D.}\ \bibnamefont
  {Teufel}}, \bibinfo {author} {\bibfnamefont {T.}~\bibnamefont {Donner}},
  \bibinfo {author} {\bibfnamefont {M.~A.}\ \bibnamefont
  {Castellanos-Beltran}}, \bibinfo {author} {\bibfnamefont {J.~W.}\
  \bibnamefont {Harlow}},\ and\ \bibinfo {author} {\bibfnamefont {K.~W.}\
  \bibnamefont {Lehnert}},\ }\bibfield  {title} {\bibinfo {title}
  {Nanomechanical motion measured with an imprecision below that at the
  standard quantum limit},\ }\href {https://doi.org/10.1038/nnano.2009.343}
  {\bibfield  {journal} {\bibinfo  {journal} {Nature Nanotechnology}\ }\textbf
  {\bibinfo {volume} {4}},\ \bibinfo {pages} {820} (\bibinfo {year}
  {2009})}\BibitemShut {NoStop}%
\bibitem [{\citenamefont {Zoepfl}\ \emph {et~al.}(2020)\citenamefont {Zoepfl},
  \citenamefont {Juan}, \citenamefont {Schneider},\ and\ \citenamefont
  {Kirchmair}}]{Zoepfl2020}%
  \BibitemOpen
  \bibfield  {author} {\bibinfo {author} {\bibfnamefont {D.}~\bibnamefont
  {Zoepfl}}, \bibinfo {author} {\bibfnamefont {M.~L.}\ \bibnamefont {Juan}},
  \bibinfo {author} {\bibfnamefont {C.~M.~F.}\ \bibnamefont {Schneider}},\ and\
  \bibinfo {author} {\bibfnamefont {G.}~\bibnamefont {Kirchmair}},\ }\bibfield
  {title} {\bibinfo {title} {Single-photon cooling in microwave
  magnetomechanics},\ }\href {https://doi.org/10.1103/PhysRevLett.125.023601}
  {\bibfield  {journal} {\bibinfo  {journal} {Phys. Rev. Lett.}\ }\textbf
  {\bibinfo {volume} {125}},\ \bibinfo {pages} {023601} (\bibinfo {year}
  {2020})}\BibitemShut {NoStop}%
\bibitem [{\citenamefont {Schmidt}\ \emph {et~al.}(2020)\citenamefont
  {Schmidt}, \citenamefont {T.~Amawi}, \citenamefont {Pogorzalek},
  \citenamefont {Deppe}, \citenamefont {Marx}, \citenamefont {Gross},\ and\
  \citenamefont {Huebl}}]{Schmidt2020}%
  \BibitemOpen
  \bibfield  {author} {\bibinfo {author} {\bibfnamefont {P.}~\bibnamefont
  {Schmidt}}, \bibinfo {author} {\bibfnamefont {M.}~\bibnamefont {T.~Amawi}},
  \bibinfo {author} {\bibfnamefont {S.}~\bibnamefont {Pogorzalek}}, \bibinfo
  {author} {\bibfnamefont {F.}~\bibnamefont {Deppe}}, \bibinfo {author}
  {\bibfnamefont {A.}~\bibnamefont {Marx}}, \bibinfo {author} {\bibfnamefont
  {R.}~\bibnamefont {Gross}},\ and\ \bibinfo {author} {\bibfnamefont
  {H.}~\bibnamefont {Huebl}},\ }\bibfield  {title} {\bibinfo {title}
  {Sideband-resolved resonator electromechanics based on a nonlinear josephson
  inductance probed on the single-photon level},\ }\href
  {https://doi.org/10.1038/s42005-020-00501-3} {\bibfield  {journal} {\bibinfo
  {journal} {Communications Physics}\ }\textbf {\bibinfo {volume} {3}},\
  \bibinfo {pages} {233} (\bibinfo {year} {2020})}\BibitemShut {NoStop}%
\bibitem [{\citenamefont {Teufel}\ \emph {et~al.}(2011)\citenamefont {Teufel},
  \citenamefont {Donner}, \citenamefont {Li}, \citenamefont {Harlow},
  \citenamefont {Allman}, \citenamefont {Cicak}, \citenamefont {Sirois},
  \citenamefont {Whittaker}, \citenamefont {Lehnert},\ and\ \citenamefont
  {Simmonds}}]{Teufel2011}%
  \BibitemOpen
  \bibfield  {author} {\bibinfo {author} {\bibfnamefont {J.~D.}\ \bibnamefont
  {Teufel}}, \bibinfo {author} {\bibfnamefont {T.}~\bibnamefont {Donner}},
  \bibinfo {author} {\bibfnamefont {D.}~\bibnamefont {Li}}, \bibinfo {author}
  {\bibfnamefont {J.~W.}\ \bibnamefont {Harlow}}, \bibinfo {author}
  {\bibfnamefont {M.~S.}\ \bibnamefont {Allman}}, \bibinfo {author}
  {\bibfnamefont {K.}~\bibnamefont {Cicak}}, \bibinfo {author} {\bibfnamefont
  {A.~J.}\ \bibnamefont {Sirois}}, \bibinfo {author} {\bibfnamefont {J.~D.}\
  \bibnamefont {Whittaker}}, \bibinfo {author} {\bibfnamefont {K.~W.}\
  \bibnamefont {Lehnert}},\ and\ \bibinfo {author} {\bibfnamefont {R.~W.}\
  \bibnamefont {Simmonds}},\ }\bibfield  {title} {\bibinfo {title} {Sideband
  cooling of micromechanical motion to the quantum ground state},\ }\href
  {https://doi.org/10.1038/nature10261} {\bibfield  {journal} {\bibinfo
  {journal} {Nature}\ }\textbf {\bibinfo {volume} {475}},\ \bibinfo {pages}
  {359} (\bibinfo {year} {2011})}\BibitemShut {NoStop}%
\bibitem [{\citenamefont {Massel}\ \emph {et~al.}(2011)\citenamefont {Massel},
  \citenamefont {Heikkil{\"a}}, \citenamefont {Pirkkalainen}, \citenamefont
  {Cho}, \citenamefont {Saloniemi}, \citenamefont {Hakonen},\ and\
  \citenamefont {Sillanp{\"a}{\"a}}}]{Massel2011}%
  \BibitemOpen
  \bibfield  {author} {\bibinfo {author} {\bibfnamefont {F.}~\bibnamefont
  {Massel}}, \bibinfo {author} {\bibfnamefont {T.~T.}\ \bibnamefont
  {Heikkil{\"a}}}, \bibinfo {author} {\bibfnamefont {J.-M.}\ \bibnamefont
  {Pirkkalainen}}, \bibinfo {author} {\bibfnamefont {S.~U.}\ \bibnamefont
  {Cho}}, \bibinfo {author} {\bibfnamefont {H.}~\bibnamefont {Saloniemi}},
  \bibinfo {author} {\bibfnamefont {P.~J.}\ \bibnamefont {Hakonen}},\ and\
  \bibinfo {author} {\bibfnamefont {M.~A.}\ \bibnamefont {Sillanp{\"a}{\"a}}},\
  }\bibfield  {title} {\bibinfo {title} {Microwave amplification with
  nanomechanical resonators},\ }\href {https://doi.org/10.1038/nature10628}
  {\bibfield  {journal} {\bibinfo  {journal} {Nature}\ }\textbf {\bibinfo
  {volume} {480}},\ \bibinfo {pages} {351} (\bibinfo {year}
  {2011})}\BibitemShut {NoStop}%
\bibitem [{\citenamefont {Suh}\ \emph {et~al.}(2013)\citenamefont {Suh},
  \citenamefont {Weinstein},\ and\ \citenamefont {Schwab}}]{Suh2013}%
  \BibitemOpen
  \bibfield  {author} {\bibinfo {author} {\bibfnamefont {J.}~\bibnamefont
  {Suh}}, \bibinfo {author} {\bibfnamefont {A.~J.}\ \bibnamefont {Weinstein}},\
  and\ \bibinfo {author} {\bibfnamefont {K.~C.}\ \bibnamefont {Schwab}},\
  }\bibfield  {title} {\bibinfo {title} {Optomechanical effects of two-level
  systems in a back-action evading measurement of micro-mechanical motion},\
  }\href {https://doi.org/10.1063/1.4816428} {\bibfield  {journal} {\bibinfo
  {journal} {Applied Physics Letters}\ }\textbf {\bibinfo {volume} {103}},\
  \bibinfo {pages} {052604} (\bibinfo {year} {2013})},\ \Eprint
  {https://arxiv.org/abs/https://doi.org/10.1063/1.4816428}
  {https://doi.org/10.1063/1.4816428} \BibitemShut {NoStop}%
\bibitem [{\citenamefont {Yuan}\ \emph {et~al.}(2015)\citenamefont {Yuan},
  \citenamefont {Singh}, \citenamefont {Blanter},\ and\ \citenamefont
  {Steele}}]{Yuan2015}%
  \BibitemOpen
  \bibfield  {author} {\bibinfo {author} {\bibfnamefont {M.}~\bibnamefont
  {Yuan}}, \bibinfo {author} {\bibfnamefont {V.}~\bibnamefont {Singh}},
  \bibinfo {author} {\bibfnamefont {Y.~M.}\ \bibnamefont {Blanter}},\ and\
  \bibinfo {author} {\bibfnamefont {G.~A.}\ \bibnamefont {Steele}},\ }\bibfield
   {title} {\bibinfo {title} {Large cooperativity and microkelvin cooling with
  a three-dimensional optomechanical cavity},\ }\href
  {https://doi.org/10.1038/ncomms9491} {\bibfield  {journal} {\bibinfo
  {journal} {Nature Communications}\ }\textbf {\bibinfo {volume} {6}},\
  \bibinfo {pages} {8491} (\bibinfo {year} {2015})}\BibitemShut {NoStop}%
\bibitem [{\citenamefont {Pirkkalainen}\ \emph {et~al.}(2015)\citenamefont
  {Pirkkalainen}, \citenamefont {Damsk\"agg}, \citenamefont {Brandt},
  \citenamefont {Massel},\ and\ \citenamefont
  {Sillanp\"a\"a}}]{Pirkkalainen2015}%
  \BibitemOpen
  \bibfield  {author} {\bibinfo {author} {\bibfnamefont {J.-M.}\ \bibnamefont
  {Pirkkalainen}}, \bibinfo {author} {\bibfnamefont {E.}~\bibnamefont
  {Damsk\"agg}}, \bibinfo {author} {\bibfnamefont {M.}~\bibnamefont {Brandt}},
  \bibinfo {author} {\bibfnamefont {F.}~\bibnamefont {Massel}},\ and\ \bibinfo
  {author} {\bibfnamefont {M.~A.}\ \bibnamefont {Sillanp\"a\"a}},\ }\bibfield
  {title} {\bibinfo {title} {Squeezing of quantum noise of motion in a
  micromechanical resonator},\ }\href
  {https://doi.org/10.1103/PhysRevLett.115.243601} {\bibfield  {journal}
  {\bibinfo  {journal} {Phys. Rev. Lett.}\ }\textbf {\bibinfo {volume} {115}},\
  \bibinfo {pages} {243601} (\bibinfo {year} {2015})}\BibitemShut {NoStop}%
\bibitem [{\citenamefont {Kalaee}\ \emph {et~al.}(2019)\citenamefont {Kalaee},
  \citenamefont {Mirhosseini}, \citenamefont {Dieterle}, \citenamefont
  {Peruzzo}, \citenamefont {Fink},\ and\ \citenamefont {Painter}}]{Kalaee2019}%
  \BibitemOpen
  \bibfield  {author} {\bibinfo {author} {\bibfnamefont {M.}~\bibnamefont
  {Kalaee}}, \bibinfo {author} {\bibfnamefont {M.}~\bibnamefont {Mirhosseini}},
  \bibinfo {author} {\bibfnamefont {P.~B.}\ \bibnamefont {Dieterle}}, \bibinfo
  {author} {\bibfnamefont {M.}~\bibnamefont {Peruzzo}}, \bibinfo {author}
  {\bibfnamefont {J.~M.}\ \bibnamefont {Fink}},\ and\ \bibinfo {author}
  {\bibfnamefont {O.}~\bibnamefont {Painter}},\ }\bibfield  {title} {\bibinfo
  {title} {Quantum electromechanics of a hypersonic crystal},\ }\href
  {https://doi.org/10.1038/s41565-019-0377-2} {\bibfield  {journal} {\bibinfo
  {journal} {Nature Nanotechnology}\ }\textbf {\bibinfo {volume} {14}},\
  \bibinfo {pages} {334} (\bibinfo {year} {2019})}\BibitemShut {NoStop}%
\bibitem [{\citenamefont {Peterson}\ \emph {et~al.}(2019)\citenamefont
  {Peterson}, \citenamefont {Kotler}, \citenamefont {Lecocq}, \citenamefont
  {Cicak}, \citenamefont {Jin}, \citenamefont {Simmonds}, \citenamefont
  {Aumentado},\ and\ \citenamefont {Teufel}}]{Peterson2019}%
  \BibitemOpen
  \bibfield  {author} {\bibinfo {author} {\bibfnamefont {G.~A.}\ \bibnamefont
  {Peterson}}, \bibinfo {author} {\bibfnamefont {S.}~\bibnamefont {Kotler}},
  \bibinfo {author} {\bibfnamefont {F.}~\bibnamefont {Lecocq}}, \bibinfo
  {author} {\bibfnamefont {K.}~\bibnamefont {Cicak}}, \bibinfo {author}
  {\bibfnamefont {X.~Y.}\ \bibnamefont {Jin}}, \bibinfo {author} {\bibfnamefont
  {R.~W.}\ \bibnamefont {Simmonds}}, \bibinfo {author} {\bibfnamefont
  {J.}~\bibnamefont {Aumentado}},\ and\ \bibinfo {author} {\bibfnamefont
  {J.~D.}\ \bibnamefont {Teufel}},\ }\bibfield  {title} {\bibinfo {title}
  {Ultrastrong parametric coupling between a superconducting cavity and a
  mechanical resonator},\ }\href
  {https://doi.org/10.1103/PhysRevLett.123.247701} {\bibfield  {journal}
  {\bibinfo  {journal} {Phys. Rev. Lett.}\ }\textbf {\bibinfo {volume} {123}},\
  \bibinfo {pages} {247701} (\bibinfo {year} {2019})}\BibitemShut {NoStop}%
\bibitem [{\citenamefont {Bothner}\ \emph {et~al.}(2020)\citenamefont
  {Bothner}, \citenamefont {Yanai}, \citenamefont {Iniguez-Rabago},
  \citenamefont {Yuan}, \citenamefont {Blanter},\ and\ \citenamefont
  {Steele}}]{Bothner2020}%
  \BibitemOpen
  \bibfield  {author} {\bibinfo {author} {\bibfnamefont {D.}~\bibnamefont
  {Bothner}}, \bibinfo {author} {\bibfnamefont {S.}~\bibnamefont {Yanai}},
  \bibinfo {author} {\bibfnamefont {A.}~\bibnamefont {Iniguez-Rabago}},
  \bibinfo {author} {\bibfnamefont {M.}~\bibnamefont {Yuan}}, \bibinfo {author}
  {\bibfnamefont {Y.~M.}\ \bibnamefont {Blanter}},\ and\ \bibinfo {author}
  {\bibfnamefont {G.~A.}\ \bibnamefont {Steele}},\ }\bibfield  {title}
  {\bibinfo {title} {Cavity electromechanics with parametric mechanical
  driving},\ }\href {https://doi.org/10.1038/s41467-020-15389-4} {\bibfield
  {journal} {\bibinfo  {journal} {Nature Communications}\ }\textbf {\bibinfo
  {volume} {11}},\ \bibinfo {pages} {1589} (\bibinfo {year}
  {2020})}\BibitemShut {NoStop}%
\bibitem [{\citenamefont {Arnold}\ \emph {et~al.}(2020)\citenamefont {Arnold},
  \citenamefont {Wulf}, \citenamefont {Barzanjeh}, \citenamefont {Redchenko},
  \citenamefont {Rueda}, \citenamefont {Hease}, \citenamefont {Hassani},\ and\
  \citenamefont {Fink}}]{Arnold2020}%
  \BibitemOpen
  \bibfield  {author} {\bibinfo {author} {\bibfnamefont {G.}~\bibnamefont
  {Arnold}}, \bibinfo {author} {\bibfnamefont {M.}~\bibnamefont {Wulf}},
  \bibinfo {author} {\bibfnamefont {S.}~\bibnamefont {Barzanjeh}}, \bibinfo
  {author} {\bibfnamefont {E.~S.}\ \bibnamefont {Redchenko}}, \bibinfo {author}
  {\bibfnamefont {A.}~\bibnamefont {Rueda}}, \bibinfo {author} {\bibfnamefont
  {W.~J.}\ \bibnamefont {Hease}}, \bibinfo {author} {\bibfnamefont
  {F.}~\bibnamefont {Hassani}},\ and\ \bibinfo {author} {\bibfnamefont {J.~M.}\
  \bibnamefont {Fink}},\ }\bibfield  {title} {\bibinfo {title} {Converting
  microwave and telecom photons with a silicon photonic nanomechanical
  interface},\ }\href {https://doi.org/10.1038/s41467-020-18269-z} {\bibfield
  {journal} {\bibinfo  {journal} {Nature Communications}\ }\textbf {\bibinfo
  {volume} {11}},\ \bibinfo {pages} {4460} (\bibinfo {year}
  {2020})}\BibitemShut {NoStop}%
\bibitem [{\citenamefont {Tinkham}(2004)}]{Tinkham2004}%
  \BibitemOpen
  \bibfield  {author} {\bibinfo {author} {\bibfnamefont {M.}~\bibnamefont
  {Tinkham}},\ }\href@noop {} {\emph {\bibinfo {title} {Introduction to
  superconductivity}}},\ \bibinfo {edition} {2nd}\ ed.\ (\bibinfo  {publisher}
  {Dover Publications},\ \bibinfo {address} {Mineola, N.Y.},\ \bibinfo {year}
  {2004})\BibitemShut {NoStop}%
\bibitem [{\citenamefont {van~der Laan}\ and\ \citenamefont
  {Ekin}(2007)}]{vanderLaan2007}%
  \BibitemOpen
  \bibfield  {author} {\bibinfo {author} {\bibfnamefont {D.~C.}\ \bibnamefont
  {van~der Laan}}\ and\ \bibinfo {author} {\bibfnamefont {J.~W.}\ \bibnamefont
  {Ekin}},\ }\bibfield  {title} {\bibinfo {title} {Large intrinsic effect of
  axial strain on the critical current of high-temperature superconductors for
  electric power applications},\ }\href {https://doi.org/10.1063/1.2435612}
  {\bibfield  {journal} {\bibinfo  {journal} {Applied Physics Letters}\
  }\textbf {\bibinfo {volume} {90}},\ \bibinfo {pages} {052506} (\bibinfo
  {year} {2007})},\ \Eprint
  {https://arxiv.org/abs/https://doi.org/10.1063/1.2435612}
  {https://doi.org/10.1063/1.2435612} \BibitemShut {NoStop}%
\bibitem [{\citenamefont {Annunziata}\ \emph {et~al.}(2010)\citenamefont
  {Annunziata}, \citenamefont {Santavicca}, \citenamefont {Frunzio},
  \citenamefont {Catelani}, \citenamefont {Rooks}, \citenamefont {Frydman},\
  and\ \citenamefont {Prober}}]{Annunziata2010}%
  \BibitemOpen
  \bibfield  {author} {\bibinfo {author} {\bibfnamefont {A.~J.}\ \bibnamefont
  {Annunziata}}, \bibinfo {author} {\bibfnamefont {D.~F.}\ \bibnamefont
  {Santavicca}}, \bibinfo {author} {\bibfnamefont {L.}~\bibnamefont {Frunzio}},
  \bibinfo {author} {\bibfnamefont {G.}~\bibnamefont {Catelani}}, \bibinfo
  {author} {\bibfnamefont {M.~J.}\ \bibnamefont {Rooks}}, \bibinfo {author}
  {\bibfnamefont {A.}~\bibnamefont {Frydman}},\ and\ \bibinfo {author}
  {\bibfnamefont {D.~E.}\ \bibnamefont {Prober}},\ }\bibfield  {title}
  {\bibinfo {title} {Tunable superconducting nanoinductors},\ }\href
  {https://doi.org/10.1088/0957-4484/21/44/445202} {\bibfield  {journal}
  {\bibinfo  {journal} {Nanotechnology}\ }\textbf {\bibinfo {volume} {21}},\
  \bibinfo {pages} {445202} (\bibinfo {year} {2010})}\BibitemShut {NoStop}%
\bibitem [{\citenamefont {Ku}\ \emph {et~al.}(2010)\citenamefont {Ku},
  \citenamefont {Manucharyan},\ and\ \citenamefont {Bezryadin}}]{Ku2010}%
  \BibitemOpen
  \bibfield  {author} {\bibinfo {author} {\bibfnamefont {J.}~\bibnamefont
  {Ku}}, \bibinfo {author} {\bibfnamefont {V.}~\bibnamefont {Manucharyan}},\
  and\ \bibinfo {author} {\bibfnamefont {A.}~\bibnamefont {Bezryadin}},\
  }\bibfield  {title} {\bibinfo {title} {Superconducting nanowires as nonlinear
  inductive elements for qubits},\ }\href
  {https://doi.org/10.1103/PhysRevB.82.134518} {\bibfield  {journal} {\bibinfo
  {journal} {Phys. Rev. B}\ }\textbf {\bibinfo {volume} {82}},\ \bibinfo
  {pages} {134518} (\bibinfo {year} {2010})}\BibitemShut {NoStop}%
\bibitem [{\citenamefont {Basset}\ \emph {et~al.}(2019)\citenamefont {Basset},
  \citenamefont {Watfa}, \citenamefont {Aiello}, \citenamefont {F\'echant},
  \citenamefont {Morvan}, \citenamefont {Estève}, \citenamefont {Gabelli},
  \citenamefont {Aprili}, \citenamefont {Weil}, \citenamefont {Kasumov},
  \citenamefont {Bouchiat},\ and\ \citenamefont {Deblock}}]{Basset2019}%
  \BibitemOpen
  \bibfield  {author} {\bibinfo {author} {\bibfnamefont {J.}~\bibnamefont
  {Basset}}, \bibinfo {author} {\bibfnamefont {D.}~\bibnamefont {Watfa}},
  \bibinfo {author} {\bibfnamefont {G.}~\bibnamefont {Aiello}}, \bibinfo
  {author} {\bibfnamefont {M.}~\bibnamefont {F\'echant}}, \bibinfo {author}
  {\bibfnamefont {A.}~\bibnamefont {Morvan}}, \bibinfo {author} {\bibfnamefont
  {J.}~\bibnamefont {Estève}}, \bibinfo {author} {\bibfnamefont
  {J.}~\bibnamefont {Gabelli}}, \bibinfo {author} {\bibfnamefont
  {M.}~\bibnamefont {Aprili}}, \bibinfo {author} {\bibfnamefont
  {R.}~\bibnamefont {Weil}}, \bibinfo {author} {\bibfnamefont {A.}~\bibnamefont
  {Kasumov}}, \bibinfo {author} {\bibfnamefont {H.}~\bibnamefont {Bouchiat}},\
  and\ \bibinfo {author} {\bibfnamefont {R.}~\bibnamefont {Deblock}},\
  }\bibfield  {title} {\bibinfo {title} {High kinetic inductance microwave
  resonators made by he-beam assisted deposition of tungsten nanowires},\
  }\href {https://doi.org/10.1063/1.5080925} {\bibfield  {journal} {\bibinfo
  {journal} {Applied Physics Letters}\ }\textbf {\bibinfo {volume} {114}},\
  \bibinfo {pages} {102601} (\bibinfo {year} {2019})},\ \Eprint
  {https://arxiv.org/abs/https://doi.org/10.1063/1.5080925}
  {https://doi.org/10.1063/1.5080925} \BibitemShut {NoStop}%
\bibitem [{\citenamefont {Thol\'en}\ \emph {et~al.}(2007)\citenamefont
  {Thol\'en}, \citenamefont {Ergül}, \citenamefont {Doherty}, \citenamefont
  {Weber}, \citenamefont {Gr\'egis},\ and\ \citenamefont
  {Haviland}}]{Tholen2007}%
  \BibitemOpen
  \bibfield  {author} {\bibinfo {author} {\bibfnamefont {E.~A.}\ \bibnamefont
  {Thol\'en}}, \bibinfo {author} {\bibfnamefont {A.}~\bibnamefont {Ergül}},
  \bibinfo {author} {\bibfnamefont {E.~M.}\ \bibnamefont {Doherty}}, \bibinfo
  {author} {\bibfnamefont {F.~M.}\ \bibnamefont {Weber}}, \bibinfo {author}
  {\bibfnamefont {F.}~\bibnamefont {Gr\'egis}},\ and\ \bibinfo {author}
  {\bibfnamefont {D.~B.}\ \bibnamefont {Haviland}},\ }\bibfield  {title}
  {\bibinfo {title} {Nonlinearities and parametric amplification in
  superconducting coplanar waveguide resonators},\ }\href
  {https://doi.org/10.1063/1.2750520} {\bibfield  {journal} {\bibinfo
  {journal} {Applied Physics Letters}\ }\textbf {\bibinfo {volume} {90}},\
  \bibinfo {pages} {253509} (\bibinfo {year} {2007})},\ \Eprint
  {https://arxiv.org/abs/https://doi.org/10.1063/1.2750520}
  {https://doi.org/10.1063/1.2750520} \BibitemShut {NoStop}%
\bibitem [{\citenamefont {Parker}\ \emph {et~al.}(2022)\citenamefont {Parker},
  \citenamefont {Savytskyi}, \citenamefont {Vine}, \citenamefont {Laucht},
  \citenamefont {Duty}, \citenamefont {Morello}, \citenamefont {Grimsmo},\ and\
  \citenamefont {Pla}}]{Parker2022}%
  \BibitemOpen
  \bibfield  {author} {\bibinfo {author} {\bibfnamefont {D.~J.}\ \bibnamefont
  {Parker}}, \bibinfo {author} {\bibfnamefont {M.}~\bibnamefont {Savytskyi}},
  \bibinfo {author} {\bibfnamefont {W.}~\bibnamefont {Vine}}, \bibinfo {author}
  {\bibfnamefont {A.}~\bibnamefont {Laucht}}, \bibinfo {author} {\bibfnamefont
  {T.}~\bibnamefont {Duty}}, \bibinfo {author} {\bibfnamefont {A.}~\bibnamefont
  {Morello}}, \bibinfo {author} {\bibfnamefont {A.~L.}\ \bibnamefont
  {Grimsmo}},\ and\ \bibinfo {author} {\bibfnamefont {J.~J.}\ \bibnamefont
  {Pla}},\ }\bibfield  {title} {\bibinfo {title} {Degenerate parametric
  amplification via three-wave mixing using kinetic inductance},\ }\href
  {https://doi.org/10.1103/PhysRevApplied.17.034064} {\bibfield  {journal}
  {\bibinfo  {journal} {Phys. Rev. Appl.}\ }\textbf {\bibinfo {volume} {17}},\
  \bibinfo {pages} {034064} (\bibinfo {year} {2022})}\BibitemShut {NoStop}%
\bibitem [{\citenamefont {Rugar}\ \emph {et~al.}(2004)\citenamefont {Rugar},
  \citenamefont {Budakian}, \citenamefont {Mamin},\ and\ \citenamefont
  {Chui}}]{Rugar2004}%
  \BibitemOpen
  \bibfield  {author} {\bibinfo {author} {\bibfnamefont {D.}~\bibnamefont
  {Rugar}}, \bibinfo {author} {\bibfnamefont {R.}~\bibnamefont {Budakian}},
  \bibinfo {author} {\bibfnamefont {H.~J.}\ \bibnamefont {Mamin}},\ and\
  \bibinfo {author} {\bibfnamefont {B.~W.}\ \bibnamefont {Chui}},\ }\bibfield
  {title} {\bibinfo {title} {Single spin detection by magnetic resonance force
  microscopy},\ }\href {https://doi.org/10.1038/nature02658} {\bibfield
  {journal} {\bibinfo  {journal} {Nature}\ }\textbf {\bibinfo {volume} {430}},\
  \bibinfo {pages} {329} (\bibinfo {year} {2004})}\BibitemShut {NoStop}%
\bibitem [{\citenamefont {Braginsky}\ \emph {et~al.}(1980)\citenamefont
  {Braginsky}, \citenamefont {Vorontsov},\ and\ \citenamefont
  {Thorne}}]{Braginsky1980}%
  \BibitemOpen
  \bibfield  {author} {\bibinfo {author} {\bibfnamefont {V.~B.}\ \bibnamefont
  {Braginsky}}, \bibinfo {author} {\bibfnamefont {Y.~I.}\ \bibnamefont
  {Vorontsov}},\ and\ \bibinfo {author} {\bibfnamefont {K.~S.}\ \bibnamefont
  {Thorne}},\ }\bibfield  {title} {\bibinfo {title} {Quantum nondemolition
  measurements},\ }\href {https://doi.org/10.1126/science.209.4456.547}
  {\bibfield  {journal} {\bibinfo  {journal} {Science}\ }\textbf {\bibinfo
  {volume} {209}},\ \bibinfo {pages} {547} (\bibinfo {year} {1980})},\ \Eprint
  {https://arxiv.org/abs/https://www.science.org/doi/pdf/10.1126/science.209.4456.547}
  {https://www.science.org/doi/pdf/10.1126/science.209.4456.547} \BibitemShut
  {NoStop}%
\bibitem [{\citenamefont {Hertzberg}\ \emph {et~al.}(2010)\citenamefont
  {Hertzberg}, \citenamefont {Rocheleau}, \citenamefont {Ndukum}, \citenamefont
  {Savva}, \citenamefont {Clerk},\ and\ \citenamefont
  {Schwab}}]{Hertzberg2010}%
  \BibitemOpen
  \bibfield  {author} {\bibinfo {author} {\bibfnamefont {J.~B.}\ \bibnamefont
  {Hertzberg}}, \bibinfo {author} {\bibfnamefont {T.}~\bibnamefont
  {Rocheleau}}, \bibinfo {author} {\bibfnamefont {T.}~\bibnamefont {Ndukum}},
  \bibinfo {author} {\bibfnamefont {M.}~\bibnamefont {Savva}}, \bibinfo
  {author} {\bibfnamefont {A.~A.}\ \bibnamefont {Clerk}},\ and\ \bibinfo
  {author} {\bibfnamefont {K.~C.}\ \bibnamefont {Schwab}},\ }\bibfield  {title}
  {\bibinfo {title} {Back-action-evading measurements of nanomechanical
  motion},\ }\href {https://doi.org/10.1038/nphys1479} {\bibfield  {journal}
  {\bibinfo  {journal} {Nature Physics}\ }\textbf {\bibinfo {volume} {6}},\
  \bibinfo {pages} {213} (\bibinfo {year} {2010})}\BibitemShut {NoStop}%
\bibitem [{\citenamefont {Suh}\ \emph {et~al.}(2014)\citenamefont {Suh},
  \citenamefont {Weinstein}, \citenamefont {Lei}, \citenamefont {Wollman},
  \citenamefont {Steinke}, \citenamefont {Meystre}, \citenamefont {Clerk},\
  and\ \citenamefont {Schwab}}]{Suh2014}%
  \BibitemOpen
  \bibfield  {author} {\bibinfo {author} {\bibfnamefont {J.}~\bibnamefont
  {Suh}}, \bibinfo {author} {\bibfnamefont {A.~J.}\ \bibnamefont {Weinstein}},
  \bibinfo {author} {\bibfnamefont {C.~U.}\ \bibnamefont {Lei}}, \bibinfo
  {author} {\bibfnamefont {E.~E.}\ \bibnamefont {Wollman}}, \bibinfo {author}
  {\bibfnamefont {S.~K.}\ \bibnamefont {Steinke}}, \bibinfo {author}
  {\bibfnamefont {P.}~\bibnamefont {Meystre}}, \bibinfo {author} {\bibfnamefont
  {A.~A.}\ \bibnamefont {Clerk}},\ and\ \bibinfo {author} {\bibfnamefont
  {K.~C.}\ \bibnamefont {Schwab}},\ }\bibfield  {title} {\bibinfo {title}
  {Mechanically detecting and avoiding the quantum fluctuations of a microwave
  field},\ }\href {https://doi.org/10.1126/science.1253258} {\bibfield
  {journal} {\bibinfo  {journal} {Science}\ }\textbf {\bibinfo {volume}
  {344}},\ \bibinfo {pages} {1262} (\bibinfo {year} {2014})},\ \Eprint
  {https://arxiv.org/abs/https://www.science.org/doi/pdf/10.1126/science.1253258}
  {https://www.science.org/doi/pdf/10.1126/science.1253258} \BibitemShut
  {NoStop}%
\bibitem [{\citenamefont {{Intermodulation Products AB}}(2023)}]{imp}%
  \BibitemOpen
  \bibfield  {author} {\bibinfo {author} {\bibnamefont {{Intermodulation
  Products AB}}},\ }\href@noop {} {\bibinfo {title} {{Intermodulation
  Products}}},\ \bibinfo {howpublished} {\url{https://intermod.pro/}} (\bibinfo
  {year} {2023}),\ \bibinfo {note} {accessed: 2023-01-19}\BibitemShut {NoStop}%
\bibitem [{\citenamefont {Roos}\ \emph {et~al.}(2023)\citenamefont {Roos},
  \citenamefont {Scarano}, \citenamefont {Arvidsson}, \citenamefont
  {Holmgren},\ and\ \citenamefont {Haviland}}]{data_repo}%
  \BibitemOpen
  \bibfield  {author} {\bibinfo {author} {\bibfnamefont {A.~K.}\ \bibnamefont
  {Roos}}, \bibinfo {author} {\bibfnamefont {E.}~\bibnamefont {Scarano}},
  \bibinfo {author} {\bibfnamefont {E.~K.}\ \bibnamefont {Arvidsson}}, \bibinfo
  {author} {\bibfnamefont {E.}~\bibnamefont {Holmgren}},\ and\ \bibinfo
  {author} {\bibfnamefont {D.~B.}\ \bibnamefont {Haviland}},\ }\href
  {https://doi.org/10.5281/zenodo.7569459} {\bibinfo {title} {Data and code for
  figures: Kinetic inductive mechano-electric transduction for nano-scale force
  sensing}} (\bibinfo {year} {2023})\BibitemShut {NoStop}%
\bibitem [{\citenamefont {{COMSOL AB, Stockholm, Sweden}}(2021)}]{comsol}%
  \BibitemOpen
  \bibfield  {author} {\bibinfo {author} {\bibnamefont {{COMSOL AB, Stockholm,
  Sweden}}},\ }\href {https://www.comsol.com} {\bibinfo {title} {Comsol
  multiphysics 6.0$ \textsuperscript{\textregistered}$}} (\bibinfo {year}
  {2021})\BibitemShut {NoStop}%
\bibitem [{\citenamefont {{Sonnet Software Inc.}}(2021)}]{sonnet}%
  \BibitemOpen
  \bibfield  {author} {\bibinfo {author} {\bibnamefont {{Sonnet Software
  Inc.}}},\ }\href {https://www.sonnetsoftware.com} {\bibinfo {title} {Sonnet}}
  (\bibinfo {year} {2021})\BibitemShut {NoStop}%
\bibitem [{\citenamefont {Samkharadze}\ \emph {et~al.}(2016)\citenamefont
  {Samkharadze}, \citenamefont {Bruno}, \citenamefont {Scarlino}, \citenamefont
  {Zheng}, \citenamefont {DiVincenzo}, \citenamefont {DiCarlo},\ and\
  \citenamefont {Vandersypen}}]{Samkharadze2016}%
  \BibitemOpen
  \bibfield  {author} {\bibinfo {author} {\bibfnamefont {N.}~\bibnamefont
  {Samkharadze}}, \bibinfo {author} {\bibfnamefont {A.}~\bibnamefont {Bruno}},
  \bibinfo {author} {\bibfnamefont {P.}~\bibnamefont {Scarlino}}, \bibinfo
  {author} {\bibfnamefont {G.}~\bibnamefont {Zheng}}, \bibinfo {author}
  {\bibfnamefont {D.~P.}\ \bibnamefont {DiVincenzo}}, \bibinfo {author}
  {\bibfnamefont {L.}~\bibnamefont {DiCarlo}},\ and\ \bibinfo {author}
  {\bibfnamefont {L.~M.~K.}\ \bibnamefont {Vandersypen}},\ }\bibfield  {title}
  {\bibinfo {title} {High-kinetic-inductance superconducting nanowire
  resonators for circuit qed in a magnetic field},\ }\href
  {https://doi.org/10.1103/PhysRevApplied.5.044004} {\bibfield  {journal}
  {\bibinfo  {journal} {Phys. Rev. Appl.}\ }\textbf {\bibinfo {volume} {5}},\
  \bibinfo {pages} {044004} (\bibinfo {year} {2016})}\BibitemShut {NoStop}%
\bibitem [{\citenamefont {Pozar}(2011)}]{Pozar2011}%
  \BibitemOpen
  \bibfield  {author} {\bibinfo {author} {\bibfnamefont {D.~M.}\ \bibnamefont
  {Pozar}},\ }\href@noop {} {\emph {\bibinfo {title} {Microwave
  Engineering}}},\ \bibinfo {edition} {4th}\ ed.\ (\bibinfo  {publisher}
  {Wiley},\ \bibinfo {address} {Hoboken, NJ},\ \bibinfo {year}
  {2011})\BibitemShut {NoStop}%
\bibitem [{\citenamefont {Fink}\ \emph {et~al.}(2016)\citenamefont {Fink},
  \citenamefont {Kalaee}, \citenamefont {Pitanti}, \citenamefont {Norte},
  \citenamefont {Heinzle}, \citenamefont {Davan{\c{c}}o}, \citenamefont
  {Srinivasan},\ and\ \citenamefont {Painter}}]{Fink2016}%
  \BibitemOpen
  \bibfield  {author} {\bibinfo {author} {\bibfnamefont {J.~M.}\ \bibnamefont
  {Fink}}, \bibinfo {author} {\bibfnamefont {M.}~\bibnamefont {Kalaee}},
  \bibinfo {author} {\bibfnamefont {A.}~\bibnamefont {Pitanti}}, \bibinfo
  {author} {\bibfnamefont {R.}~\bibnamefont {Norte}}, \bibinfo {author}
  {\bibfnamefont {L.}~\bibnamefont {Heinzle}}, \bibinfo {author} {\bibfnamefont
  {M.}~\bibnamefont {Davan{\c{c}}o}}, \bibinfo {author} {\bibfnamefont
  {K.}~\bibnamefont {Srinivasan}},\ and\ \bibinfo {author} {\bibfnamefont
  {O.}~\bibnamefont {Painter}},\ }\bibfield  {title} {\bibinfo {title} {Quantum
  electromechanics on silicon nitride nanomembranes},\ }\href
  {https://doi.org/10.1038/ncomms12396} {\bibfield  {journal} {\bibinfo
  {journal} {Nature Communications}\ }\textbf {\bibinfo {volume} {7}},\
  \bibinfo {pages} {12396} (\bibinfo {year} {2016})}\BibitemShut {NoStop}%
\bibitem [{\citenamefont {Newville}\ \emph {et~al.}(2022)\citenamefont
  {Newville}, \citenamefont {Otten}, \citenamefont {Nelson}, \citenamefont
  {Stensitzki}, \citenamefont {Ingargiola}, \citenamefont {Allan},
  \citenamefont {Fox}, \citenamefont {Carter}, \citenamefont {Michał},
  \citenamefont {Osborn}, \citenamefont {Pustakhod}, \citenamefont {lneuhaus},
  \citenamefont {Weigand}, \citenamefont {Aristov}, \citenamefont {Glenn},
  \citenamefont {Deil}, \citenamefont {Mark}, \citenamefont {Hansen},
  \citenamefont {Pasquevich}, \citenamefont {Foks}, \citenamefont {Zobrist},
  \citenamefont {Frost}, \citenamefont {Stuermer}, \citenamefont {azelcer},
  \citenamefont {Polloreno}, \citenamefont {Persaud}, \citenamefont {Nielsen},
  \citenamefont {Pompili}, \citenamefont {Caldwell},\ and\ \citenamefont
  {Hahn}}]{lmfit}%
  \BibitemOpen
  \bibfield  {author} {\bibinfo {author} {\bibfnamefont {M.}~\bibnamefont
  {Newville}}, \bibinfo {author} {\bibfnamefont {R.}~\bibnamefont {Otten}},
  \bibinfo {author} {\bibfnamefont {A.}~\bibnamefont {Nelson}}, \bibinfo
  {author} {\bibfnamefont {T.}~\bibnamefont {Stensitzki}}, \bibinfo {author}
  {\bibfnamefont {A.}~\bibnamefont {Ingargiola}}, \bibinfo {author}
  {\bibfnamefont {D.}~\bibnamefont {Allan}}, \bibinfo {author} {\bibfnamefont
  {A.}~\bibnamefont {Fox}}, \bibinfo {author} {\bibfnamefont {F.}~\bibnamefont
  {Carter}}, \bibinfo {author} {\bibnamefont {Michał}}, \bibinfo {author}
  {\bibfnamefont {R.}~\bibnamefont {Osborn}}, \bibinfo {author} {\bibfnamefont
  {D.}~\bibnamefont {Pustakhod}}, \bibinfo {author} {\bibnamefont {lneuhaus}},
  \bibinfo {author} {\bibfnamefont {S.}~\bibnamefont {Weigand}}, \bibinfo
  {author} {\bibfnamefont {A.}~\bibnamefont {Aristov}}, \bibinfo {author}
  {\bibnamefont {Glenn}}, \bibinfo {author} {\bibfnamefont {C.}~\bibnamefont
  {Deil}}, \bibinfo {author} {\bibnamefont {Mark}}, \bibinfo {author}
  {\bibfnamefont {A.~L.~R.}\ \bibnamefont {Hansen}}, \bibinfo {author}
  {\bibfnamefont {G.}~\bibnamefont {Pasquevich}}, \bibinfo {author}
  {\bibfnamefont {L.}~\bibnamefont {Foks}}, \bibinfo {author} {\bibfnamefont
  {N.}~\bibnamefont {Zobrist}}, \bibinfo {author} {\bibfnamefont
  {O.}~\bibnamefont {Frost}}, \bibinfo {author} {\bibnamefont {Stuermer}},
  \bibinfo {author} {\bibnamefont {azelcer}}, \bibinfo {author} {\bibfnamefont
  {A.}~\bibnamefont {Polloreno}}, \bibinfo {author} {\bibfnamefont
  {A.}~\bibnamefont {Persaud}}, \bibinfo {author} {\bibfnamefont {J.~H.}\
  \bibnamefont {Nielsen}}, \bibinfo {author} {\bibfnamefont {M.}~\bibnamefont
  {Pompili}}, \bibinfo {author} {\bibfnamefont {S.}~\bibnamefont {Caldwell}},\
  and\ \bibinfo {author} {\bibfnamefont {A.}~\bibnamefont {Hahn}},\ }\href
  {https://doi.org/10.5281/zenodo.7370358} {\bibinfo {title} {lmfit/lmfit-py:
  1.1.0}} (\bibinfo {year} {2022})\BibitemShut {NoStop}%
\bibitem [{\citenamefont {Golokolenov}\ \emph {et~al.}(2021)\citenamefont
  {Golokolenov}, \citenamefont {Cattiaux}, \citenamefont {Kumar}, \citenamefont
  {Sillanpää}, \citenamefont {de~L\'epinay}, \citenamefont {Fefferman},\ and\
  \citenamefont {Collin}}]{Golokolenov2021}%
  \BibitemOpen
  \bibfield  {author} {\bibinfo {author} {\bibfnamefont {I.}~\bibnamefont
  {Golokolenov}}, \bibinfo {author} {\bibfnamefont {D.}~\bibnamefont
  {Cattiaux}}, \bibinfo {author} {\bibfnamefont {S.}~\bibnamefont {Kumar}},
  \bibinfo {author} {\bibfnamefont {M.}~\bibnamefont {Sillanpää}}, \bibinfo
  {author} {\bibfnamefont {L.~M.}\ \bibnamefont {de~L\'epinay}}, \bibinfo
  {author} {\bibfnamefont {A.}~\bibnamefont {Fefferman}},\ and\ \bibinfo
  {author} {\bibfnamefont {E.}~\bibnamefont {Collin}},\ }\bibfield  {title}
  {\bibinfo {title} {Microwave single-tone optomechanics in the classical
  regime},\ }\href {https://doi.org/10.1088/1367-2630/abf983} {\bibfield
  {journal} {\bibinfo  {journal} {New Journal of Physics}\ }\textbf {\bibinfo
  {volume} {23}},\ \bibinfo {pages} {053008} (\bibinfo {year}
  {2021})}\BibitemShut {NoStop}%
\end{thebibliography}
\end{document}